\documentclass[10pt,journal,compsoc]{IEEEtran}
\ifCLASSOPTIONcompsoc
  \usepackage[nocompress]{cite}
\else
  \usepackage{cite}
\fi

\ifCLASSINFOpdf
   \usepackage[pdftex]{graphicx}
\else
\fi
\usepackage{amsmath}
\usepackage{booktabs}
\usepackage{threeparttable}
\usepackage{multirow}
\usepackage{diagbox}
\usepackage{caption}
\usepackage{subfigure}
\usepackage{multicol}
\usepackage{graphicx}

\usepackage{amsfonts}

\begin{document}

\title{DRD-Net: Detail-recovery Image Deraining via Context Aggregation Networks}

\author{\IEEEauthorblockN{Sen Deng\IEEEauthorrefmark{2},
		Mingqiang Wei\IEEEauthorrefmark{2},
		Jun Wang\IEEEauthorrefmark{2},
		Luming Liang\IEEEauthorrefmark{3},
		Haoran Xie\IEEEauthorrefmark{4},
		and~Meng~Wang\IEEEauthorrefmark{5}}
	\thanks{
		\IEEEauthorblockA{\IEEEauthorrefmark{2}S. Deng, M. Wei and J. Wang are with the School of Computer Science, Nanjing University of Aeronautics and Astronautics, Nanjing, China. }
		\IEEEauthorblockA{\IEEEauthorrefmark{3} L. Liang is with the Applied Science Group, Microsoft, Redmond, WA, 98052 USA.}
		\IEEEauthorblockA{\IEEEauthorrefmark{4}H. Xie is with the Department of Mathematics and Information Technology, Education University of Hong Kong, Hong Kong SAR, China. }
		\IEEEauthorblockA{\IEEEauthorrefmark{5}M. Wang is with the School of Computer and Information Science, Hefei University of Technology, Hefei, China. \textbf{Source code:} https://github.com/Dengsgithub/DRD-Net.}
		}}

\IEEEtitleabstractindextext{%
\begin{abstract}
Image deraining is a fundamental, yet not well-solved problem in computer vision and graphics. The traditional image deraining approaches commonly behave ineffectively in medium and heavy rain removal,
while the learning-based ones lead to image degradations such as the loss of image details, halo artifacts and/or color distortion.
Unlike existing image deraining approaches that lack the detail-recovery mechanism, we propose an end-to-end detail-recovery image deraining network (termed a DRD-Net) for single images.
We for the first time introduce two sub-networks with a comprehensive loss function which synergize to derain and recover the lost details caused by deraining.
We have three key contributions. First, we present a rain residual network to remove rain streaks from the rainy images,
which combines the \textit{squeeze-and-excitation} (SE) operation with residual blocks to make full advantage of spatial contextual information.
Second, we design a new connection style block, named \textit{structure detail context aggregation block} (SDCAB), which aggregates context feature information and has a large reception field.
Third, benefiting from the SDCAB, we construct a detail repair network to encourage the lost details to return for eliminating image degradations.
We have validated our approach on four recognized datasets (three synthetic and one real-world). Both quantitative and qualitative comparisons show that our approach outperforms
the state-of-the-art deraining methods in terms of the deraining robustness and detail accuracy. The source code has been available for public evaluation and use on GitHub.
\end{abstract}

\begin{IEEEkeywords}
DRD-Net, Image deraining, Detail-recovery, Context aggregation networks
\end{IEEEkeywords}}

\maketitle

\IEEEdisplaynontitleabstractindextext

%
\IEEEpeerreviewmaketitle

\section{Introduction}

\par Images captured in rainy days inevitably suffer from noticeable degradation of visual quality. The degradation causes significant detrimental impacts on outdoor vision-based tasks, such as video surveillance, autonomous driving, and object detection. It is, therefore, indispensable to remove rain in rainy images,
which is referred to as image deraining.

\par The ultimate goal of image deraining is to recover the ground-truth image $\textbf{B}$ from its observation $\mathbf{O = B + R}$ with the rain streaks $\textbf{R}$, which is an ill-posed problem since both the clean image and rain streaks are all unknown. Video-based deraining methods could borrow the redundant information between the sequence frames for quality rain removal \cite{rain_video_1, rain_video_2}. In contrast, single image based deraining methods ought to either draw support from priors, such as Gaussian mixture model \cite{Gaussian_mixture_model}, sparse coding \cite{sparse_coding} and low-rank representation \cite{low_rank, guo2019robust} or feed a large dataset into the well-designed deep networks \cite{detail_layer, hu2019}, due to the lack of sequence information.

\begin{figure*}[!t] \centering
	\includegraphics[width=1\linewidth]{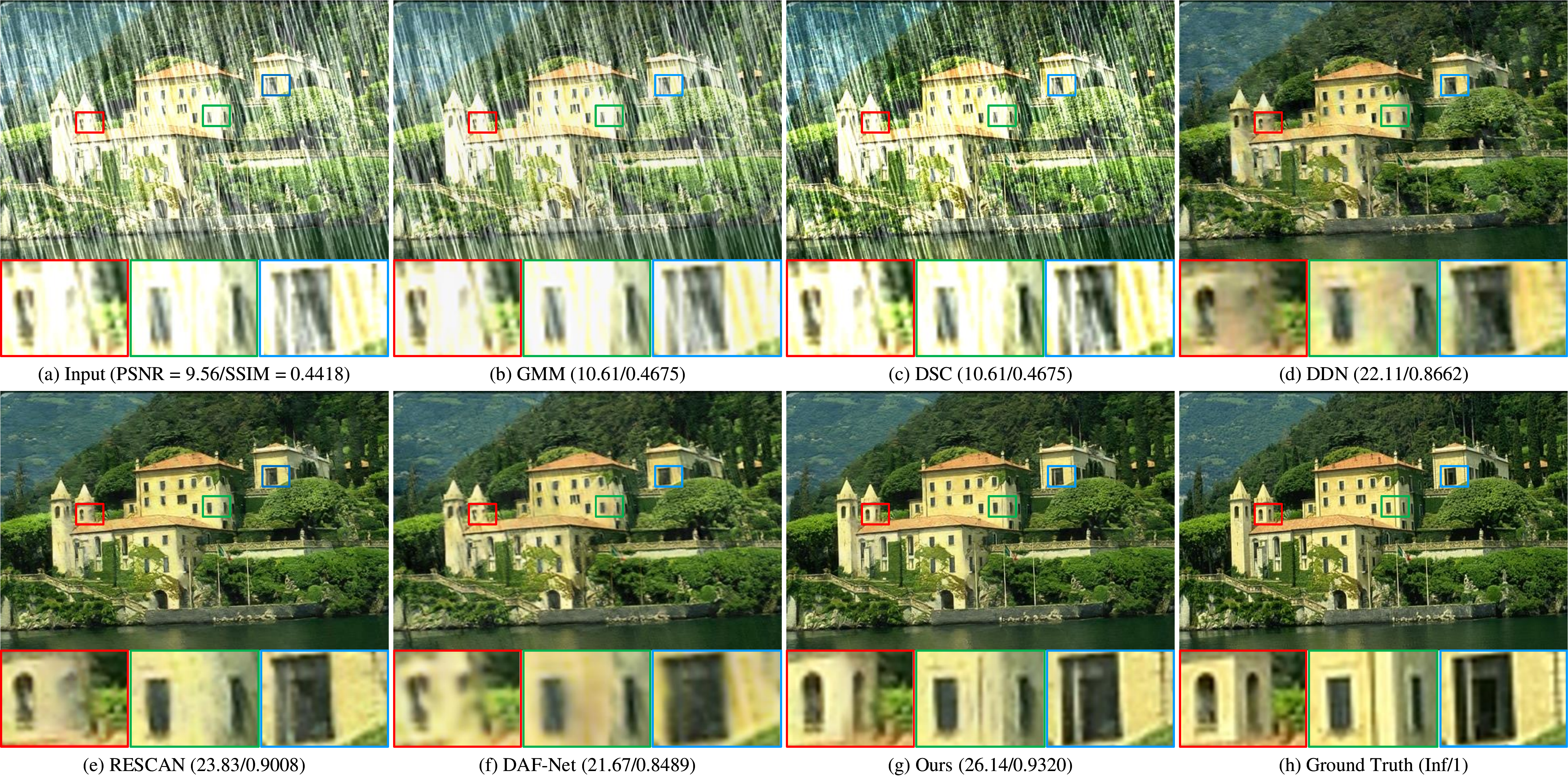}
	\caption{Image deraining results tested in the dataset of Rain200H. From (a)-(h): (a) the rainy image Castle, and the deraining results of (b) GMM \cite{Gaussian_mixture_model}, (c) DSC \cite{DSC}, (d) DDN \cite{detail_layer},
		(e) RESCAN \cite{rescan}, (f) DAF-Net \cite{hu2019}, and (g) our DRD-Net respectively, and (h) the ground-truth image. Our DRD-Net can remove the heavy rain effectively while better avoiding image degradations of the loss of image details and color distortion, due to the squeeze-and-excitation operation in the rain-removal sub-network used and an additional detail-recovery sub-network employed based on the structure detail context aggregation block. }
	\label{fig:castle}
\end{figure*}

Despite the great improvements of image deraining to produce promising deraining results when handling light-rain images, they are hindered to both remove rain streaks completely and preserve image details effectively on the images captured from the extremely bad weather. Such a phenomenon happens in Fig. \ref{fig:castle}. That is because the magnitude of image details is similar to and even smaller than that of rain streaks, but a rainy image in computer lacks semantic information to describe them separately. Therefore, the image details and rain streaks are commonly removed simultaneously. No state-of-the-art methods can serve as an image deraining panacea for various applications: they produce the deraining results with a tradeoff between rain removal and image detail maintenance.

Different from existing image deraining methods that attempt to maintain image details, we look at this intriguing question: now that image deraining inevitably leads to image detail blurring in nature, are single images with their details lost during image deraining, reversible to their artifact-free status?  

We for the first time propose an end-to-end detail-recovery image deraining network (DRD-Net) based on the context aggregation networks, which can remove rain streaks completely while recovering the original image details clearly. The success of our DRD-Net attributes to the top three factors. 1) For complete rain removal, we design a \textit{squeeze-and-excitation} (SE) based rain residual network, which can aggregate feature maps in the same convolutional layer to make full advantage of spatial contextual information. 2) For detail recovery, we build an additional detail repair network to encourage the lost details to return to the derained image. 3) We propose the \textit{structure detail context aggregation block} (SDCAB), which has larger reception fields and makes full use of the rain-free image patches for better detail recovery.
To the best of our knowledge, there is no similar image deraining architecture that can recover the image details once they are lost during deraining.

Qualitative and quantitative assessments show that, our method outperforms the state-of-the-art learning-based and traditional approaches in terms of handling inputs from both synthetic and real-world datasets.

\section{Related Work}

\par Existing rain removal methods are video-based and single image-based. Without the temporal information, single image-based methods are more challenging than video-based methods. We briefly review the related methods as follows.

\subsection{Video-based Methods}
\par Owing to the redundant information of the sequence frames in videos, rain streaks can be more easily identified and removed \cite{rain_video_1, rain_video_2, rain_video_3}. \cite{rain_video_1} replaces the intensity of a rainy pixel by averaging the corresponding pixel's intensities in the adjacent frames. \cite{rain_video_4} detects the rain streaks based on the histogram of rain streaks orientations. \cite{rain_video_5} summarizes the video-based deraining methods that have been proposed in recent years.

\subsection{Single Image-based Methods}
\par For rain removal from single images, existing methods fall into two categories: the traditional methods and the deep-learning based methods.

\par \textbf{Traditional Methods:} Various image priors have been proposed to remove rain from single images. They assume that rain streaks \textbf{R} are sparse and in similar directions. Under this assumption, they decompose the input image \textbf{O} into the rain-free background scene \textbf{B} and the rain streaks layer \textbf{R}. \cite{sparse_code} separates the rain streaks from high frequency using dictionary learning. \cite{sparse_coding} presents a discriminative sparse coding for separating rain streaks from the background image based on image patches. In \cite{Gaussian_mixture_model}, Gaussian mixture models (GMM), as a prior, is proposed to decompose the input image into the rain streaks and the background layer. \cite{zhu_lei} first detects rain-dominant regions and then the detected regions are utilized as a guidance image to help separate rain streaks from the background layer. \cite{low_rank} leverages the low-rank property of rain streaks to separate the two layers.

\par \textbf{Deep Learning-based Methods:} Deep-learning based methods have been introduced to single image deraining by \cite{detail_layer}, which boost the performance significantly. The authors decompose the input image into the low and high-frequency layers, and then build a deep residual network which learns a function mapping the high frequency parts to rain streaks. Real rain streaks have irregular distribution, since raindrops in the air have various appearances and occur at different distances from the camera. Therefore, a new rain model has been formulated as
\begin{equation}
\mathbf{O = B + }\sum\limits_{i=1}^{n} \mathbf{R}^{i},
\end{equation}
where $n$ is the number of rain streaks layers, $R^{i}$ represents the $i$-th rain streaks layer with the same direction. Furthermore, the real rain streaks can be more complicated, especially in the case of heavy rain. The rain appearance is also formed by the accumulation of rain streaks, which is similar to mist or fog \cite{shuju}. In order to imitate the real rainy environment, they propose an improved rain model as
\begin{equation}
\mathbf{O} = \alpha \left(\mathbf{B} + \sum\limits_{i=1}^{n} \mathbf{R}^{i} \right) + \left( 1 - \alpha \right) \textbf{A},
\end{equation}
where $\alpha$ is the atmospheric transmission, \textbf{A} is the global atmospheric light \cite{shuju}.

Later, \cite{zhang_gan_dl} presents a conditional generative adversarial network (GAN) and use the perceptual loss to refine the results. \cite{Yang_dl} develops a deep recurrent dilated joint rain streaks detection and removal network to remove the rain streaks. \cite{rescan} proposes the multi-stage networks based on the recurrent neural network architecture to remove rain streaks in different directions. \cite{zhang_density_dl} presents a density-aware multi-stream connected network for deraining. By maintaining negative residual features, \cite{fan_residual_dl} builds a residual-guided network for removing the rain streaks from single images.

\begin{figure*}[!t] \centering
	\includegraphics[width=1\linewidth]{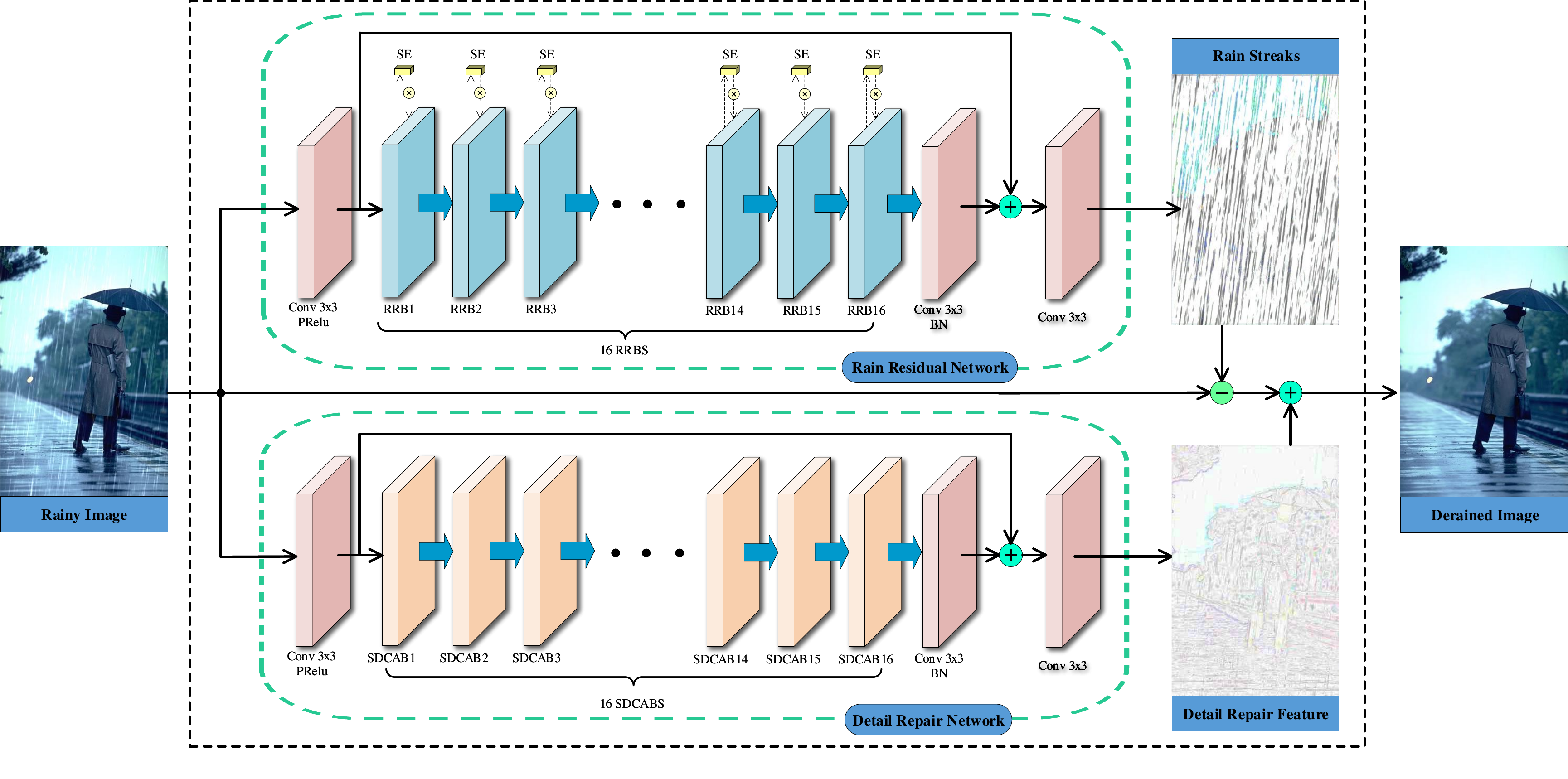}
	\caption{DRD-Net consists of two-sub networks, i.e., the rain removal network and the detail repair network. The first sub-network, which combines the squeeze-and-excitation (SE) operation with residual blocks to make full advantage of spatial contextual information, aims at removing rain streaks from the rainy image. And the second sub-network, which integrates the structure detail context aggregation block (SDCAB) to aggregate context feature information for a large reception field, seeks to recover the lost details to the derained image.}
	\label{fig:network framework}
\end{figure*}

\section{Overview}


\par Image deraining inevitably leads to image detail blurring, because rain streaks and image details are all of high-frequency information in nature. Unfortunately, existing approaches lack the mechanism to recover the image details once they are lost during image deraining. For both rain removal and detail recovery of single images, we propose two sub-networks which work together as shown in Fig. \ref{fig:network framework}. On one hand, we introduce a rain residual network to train a function that maps the rainy images to their rain streaks. Therefore, we can obtain the preliminarily derained images by separating the rain streaks from the rainy images. On the other hand, different from other methods which try to decompose a single rainy image into a background layer and a rain streaks layer, we present an additional detail repair network to find back the lost details.

\begin{figure*}[!t] \centering
	\includegraphics[width=1\linewidth]{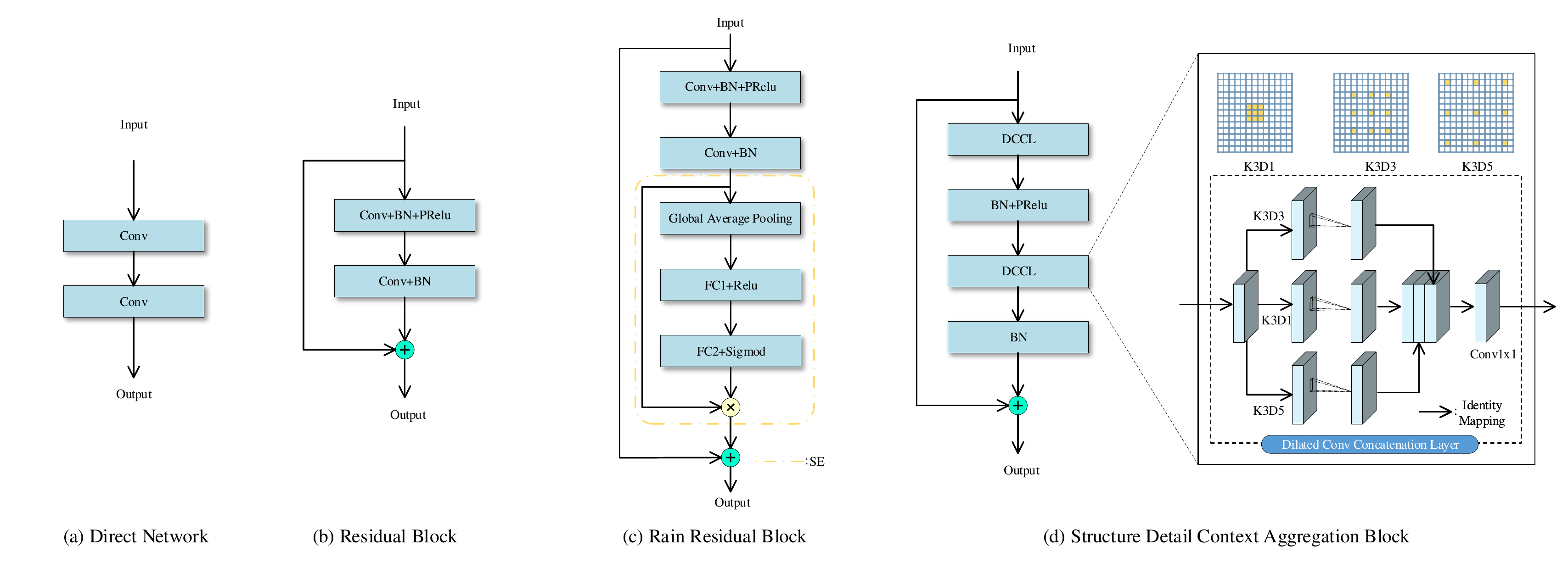}
	\caption{Illustration of different convolution styles. From (a)-(d): (a) direct network, (b) residual block, (c) rain residual block with SE used in our rain residual network, and (d) structure detail context aggregation block used in our detail repair network.}
	\label{fig:different connect}
\end{figure*}

\section{Detail-recovery Image Deraining Network (DRD-Net)}

\par DRD-Net consists of two sub-networks with a comprehensive loss function which synergize to derain and recover the
lost details caused by deraining. In the following, we will introduce the rain residual network and the detail repair network, respectively.

\subsection{Rain Residual Network}

\par Based on the observation that the rain streaks $\mathbf{R}$ are sparser than the rain-free background scene $\mathbf{B}$ \cite{rescan}, we learn a function by training a residual network, which maps the rainy image $\mathbf{O}$ to rain streaks $\mathbf{R}$. We train such a network by minimizing the loss function as
\begin{equation}
Loss_{r} = \sum\limits_{i \in N(D)}||f(\mathbf{O}_{i})-\hat{R_{i}}||^{2},
\end{equation}
where the $f(\cdot)$ is the function that we try to learn, $\mathbf{O}_{i}$ is a rainy image and $\hat{R_{i}}$ is the ground-truth rain streak layer in the training dataset $D$. $N(D)$ denotes the number of images in $D$.

\par The architecture of our rain residual network is shown in the upper part of Fig. \ref{fig:network framework}, which utilizes the Squeeze-and-Excitation (SE) \cite{SE} into the residual block. The rain residual network includes 3 convolution layers and 16 rain residual blocks. The first layer can be interpreted as an encoder, which is used to transform the rainy image into the feature maps, and the last two layers are used to recover the RGB channels from feature maps.

\par Mathematically, the rain residual block can be formulated as
\begin{equation}
RRB = SE(Res(\mathbf{X_{0}})),
\end{equation}
where $RRB$ is the output of the rain residual block, $SE(\cdot)$ and $Res(\cdot)$ denote the squeeze-and-excitation operation and the residual block shown in Fig. \ref{fig:different connect}(c) respectively, and $\mathbf{X_{0}}$ is the input signal.
\par By removing the image indexing, the rain residual network structure can be expressed as
\begin{equation}
\begin{split}
layer_{0} & = PRelu(Conv_{3 \times 3}(\mathbf{O})), \\
layer_{i} & = RRB(layer_{i-1})), \\
layer_{d-1} & = BN(Conv_{3 \times 3}(layer_{d-2})), \\
\mathbf{R} & =  Conv_{3 \times 3}(Add[layer_{0}, layer_{d-1}]),
\end{split}
\end{equation}
where d is the total number of layers, $BN(\cdot)$ denotes the batch normalization, which is used to reduce internal covariate shift \cite{BN}. $PRelu(\cdot)$ indicates a non-linearity activation function \cite{PRelu} and $Conv_{3 \times 3}$ is the convolution operation with the kernel size being equal to 3.

\begin{figure}[!t] \centering
	\includegraphics[width=1\linewidth]{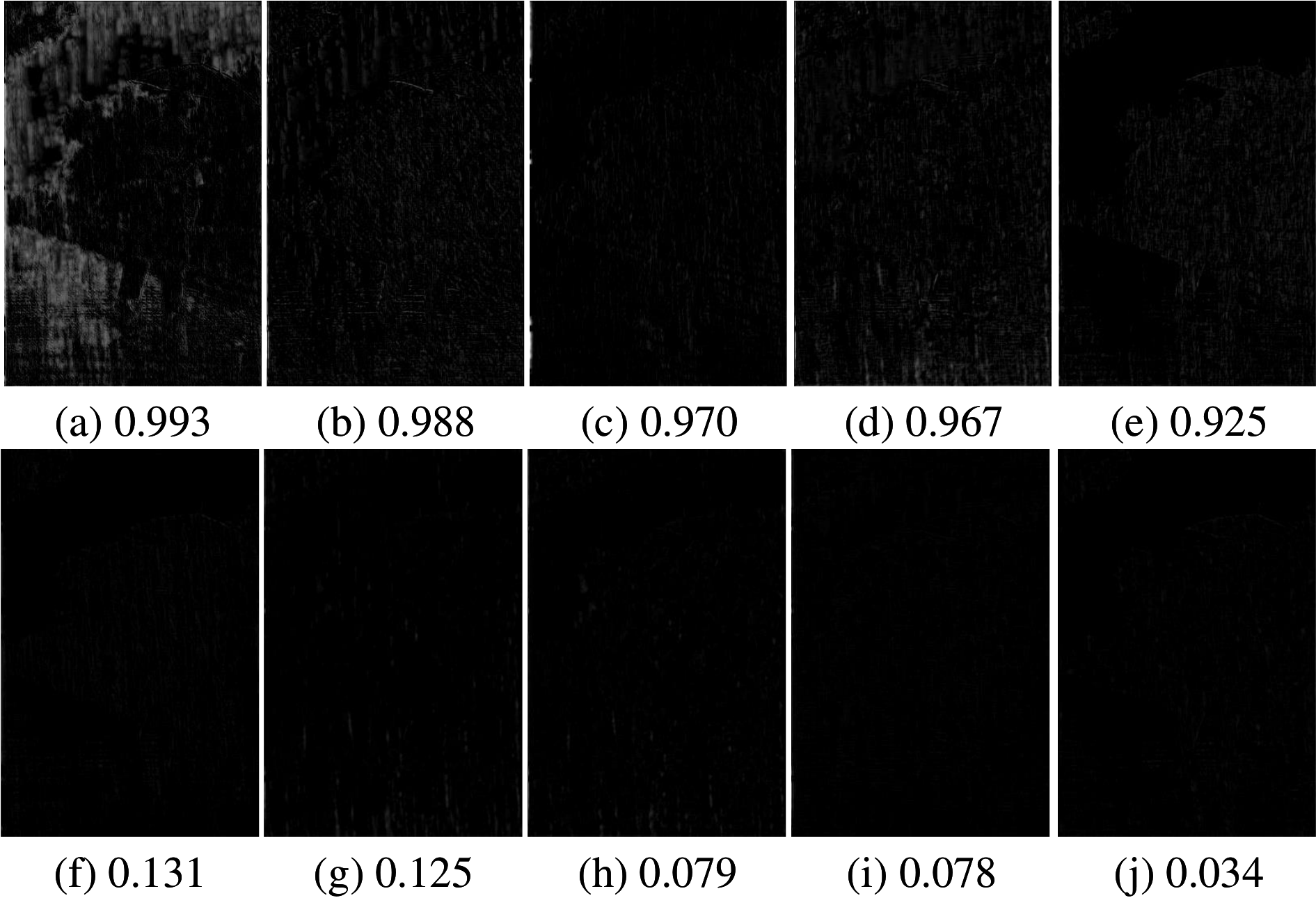}
	\caption{Feature maps with different weights. The images in (a)-(e) denote the top five high weighted feature maps, and the images in (f)-(i) denote the top five low weighted feature maps.}
	\label{fig:se}
\end{figure}

\textbf{Remark.} Spatial contextual information has proved to be effective in single image deraining \cite{Huang2012}, \cite{context_prove}. Nevertheless, the different feature channels in the same layer are independent and have little correlation during the traditional convolution operation. A main difference from the the common residual block is that we combine the SE operation into the residual block in our network. Since SE can model a correlation between different feature channels, we can intensify the feature channel which has more context information by giving a larger weight. Conversely, the feature channels that have less spatial contextual information will just receive a small weight. All the weights of different channels are learned by the rain residual network automatically during the training steps. To explore the correlation between the SE weight and the content of layers, we show the top five high/low-weighted feature maps in Fig. \ref{fig:se}. One can observe that the feature maps with more spatial contextual information have received a higher weight as expected.

\subsection{Detail Repair Network}
\par Now that image deraining inevitably leads to image degradations of the loss of image details, halo artifacts and/or color distortion in nature, we can train additional geometry-recovery network that makes the detail-lost images be reversible to their artifact-free status. Based on the preliminarily derained image $\mathbf{I}_{p}$ which is obtained by subtracting the rain streaks $\mathbf{R}$ from the rainy image $\mathbf{O}$, we can train a function to encourage the lost details to return by optimizing the loss function as
\begin{equation}
Loss_{d} = \sum\limits_{i \in N(D)}||(\mathbf{I}_{p, i} + g(\mathbf{O}_{i}))-\hat{I_{i}}||^{2},
\end{equation}
where the $g(\cdot)$ is the function that we try to learn, $\mathbf{O}_{i}$ is a rainy image.  $\hat{I_{i}}$ is the ground-truth rain-free image in the training dataset $D$. $N(D)$ denotes the number of images in $D$.
\begin{figure*}[!t] \centering
	\includegraphics[width=1\linewidth]{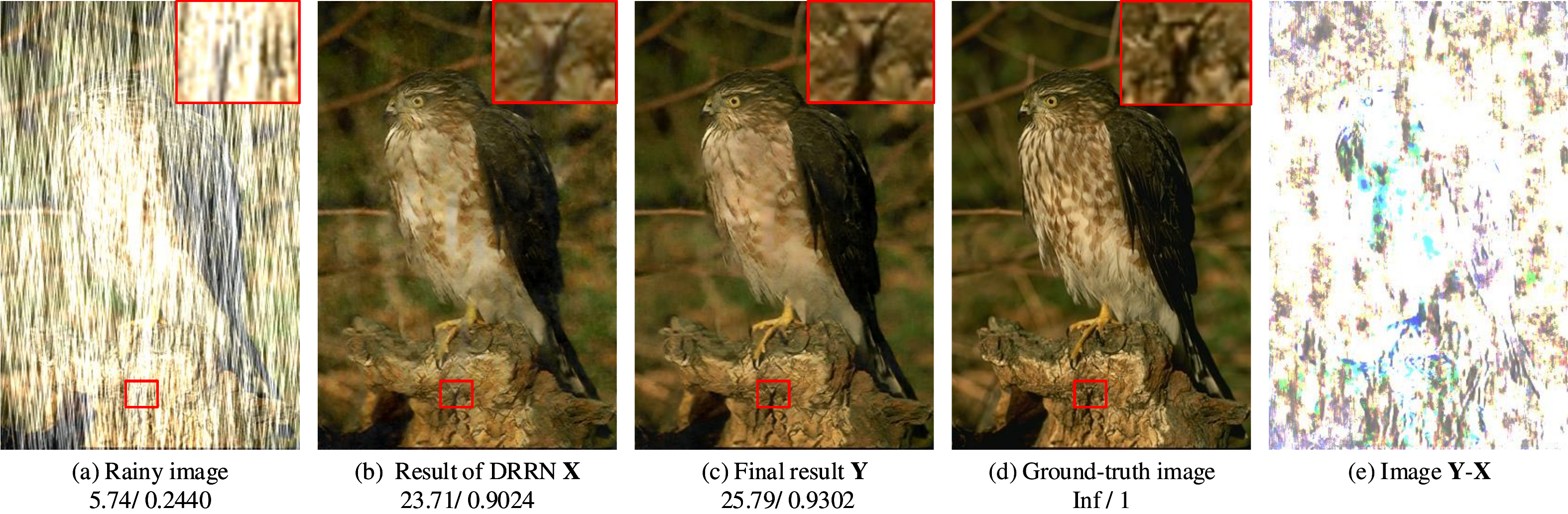}
	\caption{Image deraining results tested in the dataset of Rain200L. From (a)-(e): (a) the input rainy image, (b) the result \textbf{X} by only using the rain residual network (i.e., without the detail repair network), (c) the result \textbf{Y} by the DRD-Net, (d) the ground-truth image, and (e) the image of \textbf{Y-X} (note: we have inverted the image \textbf{Y-X} for better visualization).}
	\label{fig:Y-X}
\end{figure*}
\begin{table*}[htbp]
	\def\arraystretch{1.2}
	\centering	
	\caption{The detailed architecture of the detail repair network. }
	\label{Tab:struct}
	\resizebox{\textwidth}{!}{
		\setlength{\tabcolsep}{4mm}{
			\begin{tabular}{lccccccccc}
				\hline
				\textbf{Layer} & 0 & 1 & 2  & $\dots$ & d  & $\dots$ & 16 & 17  &18  \\ \hline
				\textit{Convolution}      & $3 \times 3$    & $3 \times 3$     & $3 \times 3$     & $\dots$    & $3 \times 3$   & $\dots$     & $3 \times 3$    & $3 \times 3$    & $3 \times 3$   \\
				\textit{SDCAB}      & No    & Yes     & Yes     & $\dots$    & Yes   & $\dots$    & Yes   & No    & No   \\
				\textit{Dilation}         & 1         & 7       & 7  & $\dots$  & 7  & $\dots$    & 7         & 1   & 1\\
				\textit{Receptive field}  & $3 \times 3$    & $17 \times 17$    & $31 \times 31$   & $\dots$  & $(d-1) \times 14+17$  & $\dots$     & $227 \times 227$       & $229 \times 229$         & $231 \times 231$   \\ \hline
			\end{tabular}
		}
	}
\end{table*}
\par We design a novel connection style block, named structure detail context aggregation block (SDCAB), which aggregates context feature information and has a large reception field. The SDCAB consists of different scales of dilation convolutions and $1 \times 1$ convolutions as shown in Fig. \ref{fig:different connect}(d). Since a large receptive field is very helpful to acquire much contextual information \cite{rescan}, we present 3 dilated convolutions whose dilation scales are 1, 3 and 5 in the SDCAB. Then, in order to extract the most important features, we concatenate the output of dilated convolutions and utilize the $1 \times 1$ convolution to reduce the feature dimensions. For reducing the complexity in training, the residual network is also introduced into the SDCAB.

\par From Fig. \ref{fig:different connect}(d), the dilated convolution concatenation layer (DCCL) can be expressed as
\begin{equation}
\begin{split}
DCCL = Conv_{1 \times 1}(Cat[Conv_{3\times 3, d_{1}}(X), \\
Conv_{3\times 3, d_{3}}(X), Conv_{3\times 3, d_{5}}(X)]),
\end{split}
\end{equation}
where $Conv_{x \times x, d_{y}}$ denotes the dilated convolutions with the kernel size of $i \times i$, and the dilation scale is $y$. $Cat(\cdot)$ is a concatenating operation and the $X$ is the input feature.
\par Mathematically, the SDCAB can be formulated as
\begin{equation}
SDCAB = Add[X_{input}, BN(DCCL_{2})],
\end{equation}
where $DCCL_{2}$ is described as
\begin{equation}
DCCL_{2} = PRelu(BN(DCCL_{1}(X_{input}))),
\end{equation}
\par Like the rain residual network, the repair network structure can be expressed as
\begin{equation}
\begin{split}
layer_{0} & = PRelu(Conv_{3 \times 3}(\mathbf{O})), \\
layer_{i} & = SDCAB(layer_{i-1})), \\
layer_{d-1} & = BN(Conv_{3 \times 3}(layer_{d-2})), \\
\mathbf{I}_{r} & =  Conv_{3 \times 3}(Add[layer_{0}, layer_{d-1}]),\\
\mathbf{I}_{c} & = Add[(\mathbf{I}_{p}, \mathbf{I}_{r})],
\end{split}
\end{equation}
where $\mathbf{I}_{r}$ is the image of found details and $\mathbf{I}_{c}$ is the final rain-free and geometry-recovery image.

\begin{figure*}[ht] \centering
	\includegraphics[width=1\linewidth]{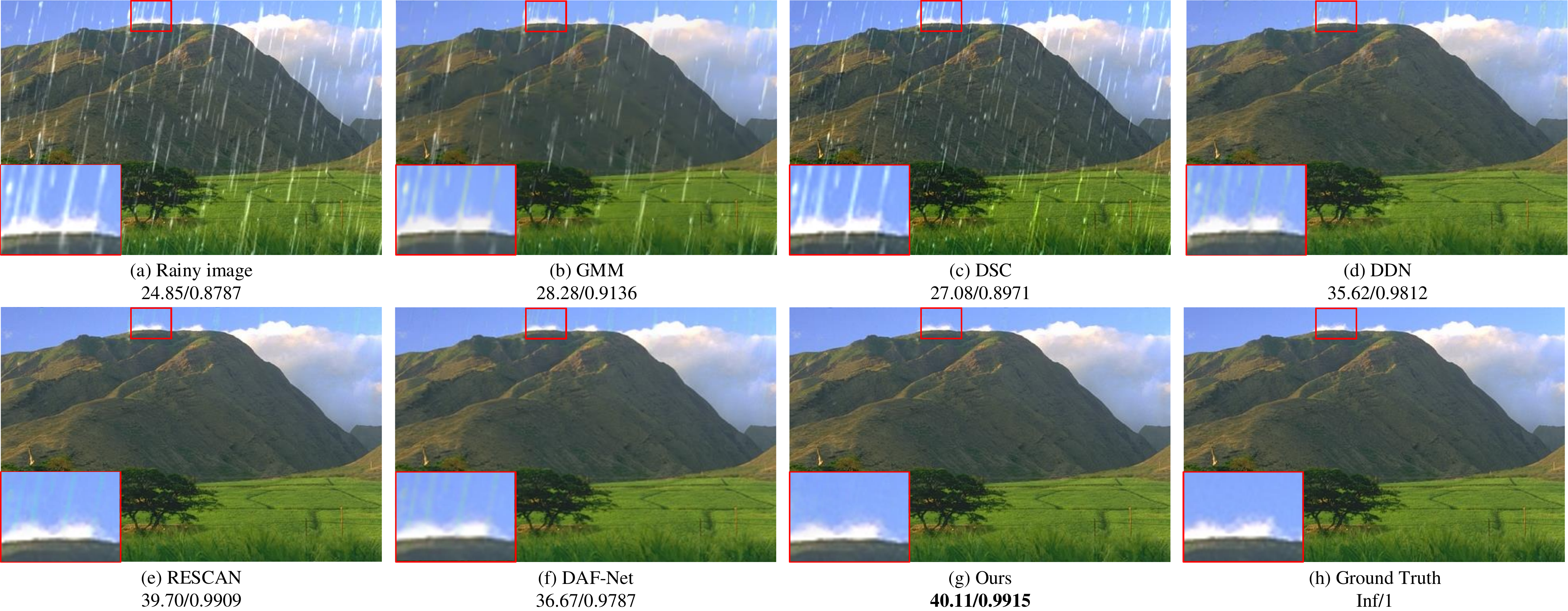}
	\caption{Image deraining results tested in the dataset of Rain200L. From (a)-(h): (a) the rainy image Mountain, and the deraining results of (b) GMM \cite{Gaussian_mixture_model}, (c) DSC \cite{DSC}, (d) DDN \cite{detail_layer},
		(e) RESCAN \cite{rescan}, (f) DAF-Net \cite{hu2019}, and (g) our DRD-Net respectively, and (h) the ground-truth image.}
	\label{fig:mountain}
\end{figure*}

\begin{figure*}[!t] \centering
	\includegraphics[width=1\linewidth]{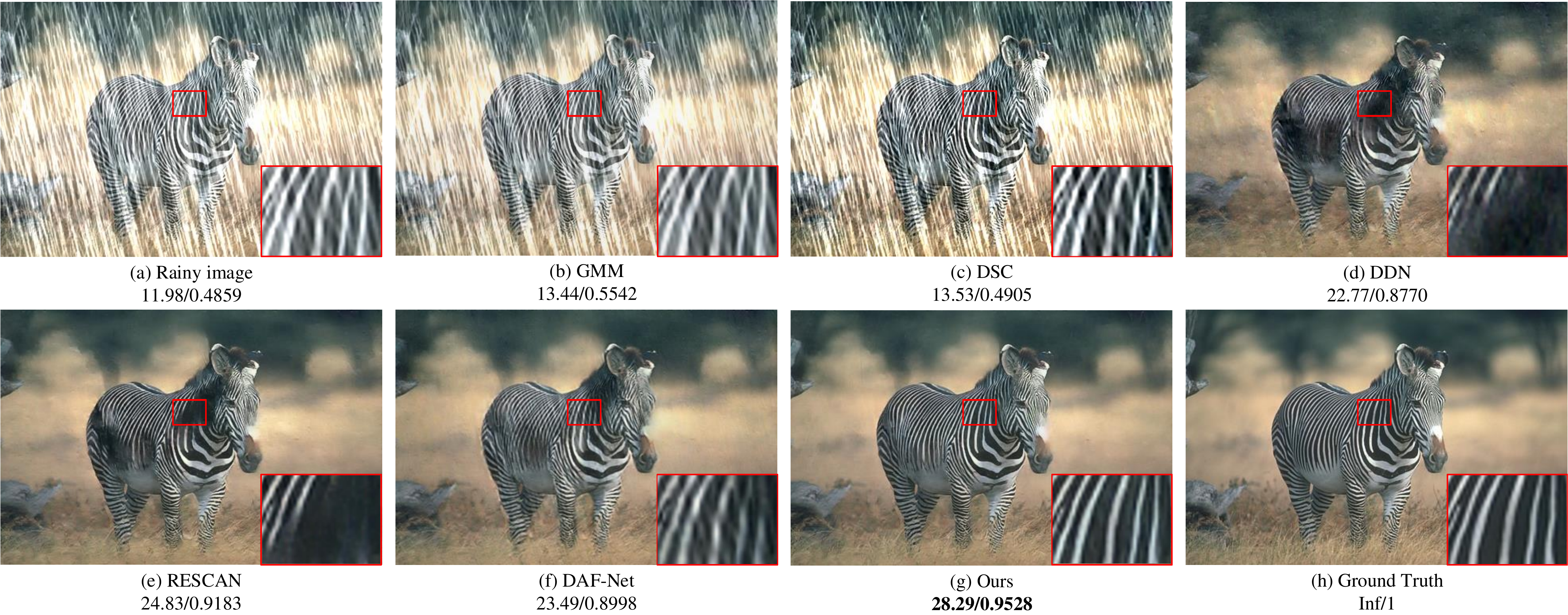}
	\caption{Image deraining results tested in the dataset of Rain200H. From (a)-(h): (a) the rainy image Zebra, and the deraining results of (b) GMM \cite{Gaussian_mixture_model}, (c) DSC \cite{DSC}, (d) DDN \cite{detail_layer},
		(e) RESCAN \cite{rescan}, (f) DAF-Net \cite{hu2019}, and (g) our DRD-Net respectively, and (h) the ground-truth image.}
	\label{fig:zebra}
\end{figure*}

\textbf{Remark.}  A large receptive field plays an important role in obtaining more information. With a larger receptive field, we can obtain more context information, which is helpful to find back the lost details. The detail of the detail repair network is outlined in Tabel \ref{Tab:struct}, where the network has a large receptive field. We can observe from Fig. \ref{fig:Y-X} that, the DRD-Net has found back the details to the image $Y$ that were lost by filtering the rainy image to obtain $X$. We have provided more experiments on three synthetic datasets to compare the performance of image deraining with and without the additional detail repair network (DRN) in Table \ref{Tab:ablation_1}, from which one can observe that our DRD-Net outperforms the single rain removal network, which implies that the detail repair network can truly find back the lost details.
\subsection{Comprehensive Loss Function}

\par As mentioned above, we employ the simplest $L_{2}$ loss as our objective function. The comprehensive loss function of our two-sub networks can be formulated as
\begin{equation}
\begin{split}
Loss_{total} & =  \lambda_{1} \sum_{i\in N(D)}||f(\mathbf{O}_{i})-\hat{R}_{i}||^{2} \\
& +  \lambda_{2}\sum_{i\in N(D)}||(\mathbf{I}_{p, i} + g(\mathbf{O}_{i}))-\hat{I}_{i}||^{2},
\end{split}
\end{equation}
where $\mathbf{O}_{i}$ denotes the i-th input rainy image, $\mathbf{I}_{p,i}$ denotes the preliminarily derained image which is obtained by subtracting the rain streaks $\mathbf{R}_{i}$ from $\mathbf{O}_{i}$, $\hat{R}_{i}$ and $\hat{I}_{i}$ are the rain-streaks image and the rain-free image respectively, $ \lambda_{1}$ and $ \lambda_{2}$ are two parameters to balance the two sub-loss functions, which in our experiments are fixed to be $0.1$ and $1.0$ respectively.

\section{Experiment and Discussions}

\par In this section, we present the details of our experimental settings. To validate our DRD-Net, we have compared it with several state-of-the-art methods on three two synthetic datasets and a real-world dataset.
\begin{figure*}[ht] \centering
	\includegraphics[width=1\linewidth]{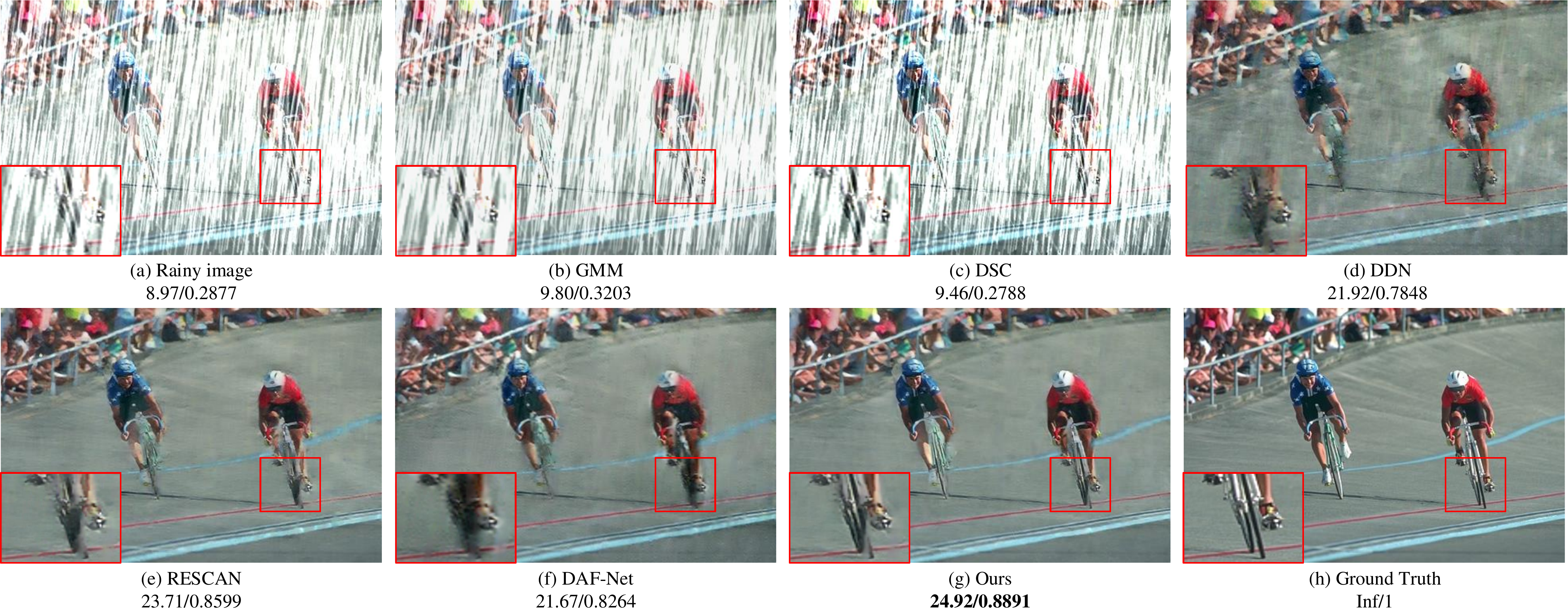}
	\caption{Image deraining results tested in the dataset of Rain200H. From (a)-(h): (a) the rainy image Match, and the deraining results of (b) GMM \cite{Gaussian_mixture_model}, (c) DSC \cite{DSC}, (d) DDN \cite{detail_layer},
		(e) RESCAN \cite{rescan}, (f) DAF-Net \cite{hu2019}, and (g) our DRD-Net respectively, and (h) the ground-truth image.}
	\label{fig:match}
\end{figure*}

\begin{table*}[htbp]
	\centering
	\caption{Quantitative experiments evaluated on three recognized synthetic datasets. The best and the second best results have been boldfaced and underlined.}
	\label{Tab:compare}
	\begin{threeparttable}
		\centering
		\setlength{\tabcolsep}{7mm}{
			\begin{tabular}{ccccccc}
				\toprule
				\multirow{2}{*}{Dataset}&
				\multicolumn{2}{c}{Rain200L}&\multicolumn{2}{c}{Rain200H}&\multicolumn{2}{c}{Rain800}\cr
				\cmidrule(lr){2-3} \cmidrule(lr){4-5} \cmidrule(lr){6-7}
				& PSNR & SSIM & PSNR & SSIM & PSNR & SSIM \cr
				\midrule
				GMM \cite{Gaussian_mixture_model}    & 27.16 & 0.8982 & 13.04 & 0.4673 & 24.04 & 0.8675 \cr
				DSC \cite{DSC}      & 25.68 & 0.8751 & 13.17 & 0.4272 & 20.95 & 0.7530 \cr
				DDN \cite{detail_layer} & 33.01 & 0.9692 & 24.64 & 0.8489 & 24.04 & 0.8675 \cr
				RESCAN \cite{rescan}    & \underline{37.07} & \underline{0.9867} & \underline{26.60} & \underline{0.8974} & 24.09 & 0.8410 \cr
				DAF-Net \cite{hu2019} & 32.07 & 0.9641 & 24.65 & 0.8607 & \underline{25.27} & \underline{0.8895} \cr
				Ours    & \textbf{37.15} & \textbf{0.9873} & \textbf{28.16} & \textbf{0.9201} & \textbf{26.32}& \textbf{0.9018} \cr
				\bottomrule
			\end{tabular}
		}
	\end{threeparttable}
\end{table*}

\subsection{Experiment Settings}
\par \quad \textbf{Synthetic Datasets:} On account of the difficulty in acquiring the rainy/clean image pair datasets, we use the synthetic datasets to train our network. \cite{zhang_gan_dl} provides a synthetic dataset named Rain800, which contains 700 training images and 100 testing image. \cite{shuju} collects and synthesizes 2 datasets, including Rain200L and Rain200H. Both Rain200L and Rain200H consist of 1800 training images and 200 test images.

\par \textbf{Real-world Datasets:} \cite{zhang_gan_dl} and \cite{shuju} also supply some real-word rainy images to validate the robustness of deraining methods. we use those images for objective evaluation.

\par \textbf{Training Details:} We set the total number of epochs for training to be 120, and each epoch includes 1000 iterations. During the training process, we set the depth of our network to be 35, and utilize the non-linear activation PRelu \cite{PRelu}. For optimizing our network, the Adam \cite{adam} is adopted with a min-batch size of 4 to train the network. We initialize the learning rate as 0.01, which is divided by 2 every 15 epochs. All the experiments are performed by using an Nvidia 1080Ti GPU.

\par \textbf{Quality Comparisons:} To quantitatively compare the performance of different methods, we adopt the Peak Signal to Noise Ratio (PSNR) \cite{psnr} and the Structure Similarity Index (SSIM) \cite{ssim} to evaluate the quality of single images deraining. Because of lacking the ground-truth images of real-world datasets, we only evaluate the performance on the real-word datasets visually.

\subsection{Test on Synthetic Datasets}

\par We quantitatively compare our method with several state-of-the-art deraining methods in this section, including 2 traditional methods, i.e., GMM \cite{Gaussian_mixture_model}, and DSC \cite{DSC}, and 4 learning-based methods, i.e., DDN \cite{detail_layer}, RESCAN \cite{rescan}, UGSM \cite{AMM}, and DAF-Net \cite{hu2019}. All these methods are performed in the same training and testing datasets for fair comparison. The comparison results are shown in Table \ref{Tab:compare}. As shown in Table \ref{Tab:compare}, our DRD-Net obtains the higher values of PSNR and SSIM than the other methods on those three datasets.

\par Our DRD-Net can effectively avoid image degradations caused by the deraining process, which can be demonstrated in Fig. \ref{fig:mountain}. As shown in these results, although most approaches can remove the rain streaks from the rainy image, the halo artifacts and color distortion have appeared after deraining. By contrast, our method can effectively deal with this problem. 

Moreover, it is challenging for most approaches to maintain/recover the details from heavy rainy images as shown in both Figs. \ref{fig:zebra} and \ref{fig:match}. The white stripes of the zebra in Fig. \ref{fig:zebra} and the bicycle in Fig. \ref{fig:match} are blurred severely by the compared approaches while they are considered as image details and recovered well by our DRD-Net.


\begin{figure*}[!t] \centering
	\includegraphics[width=1\linewidth]{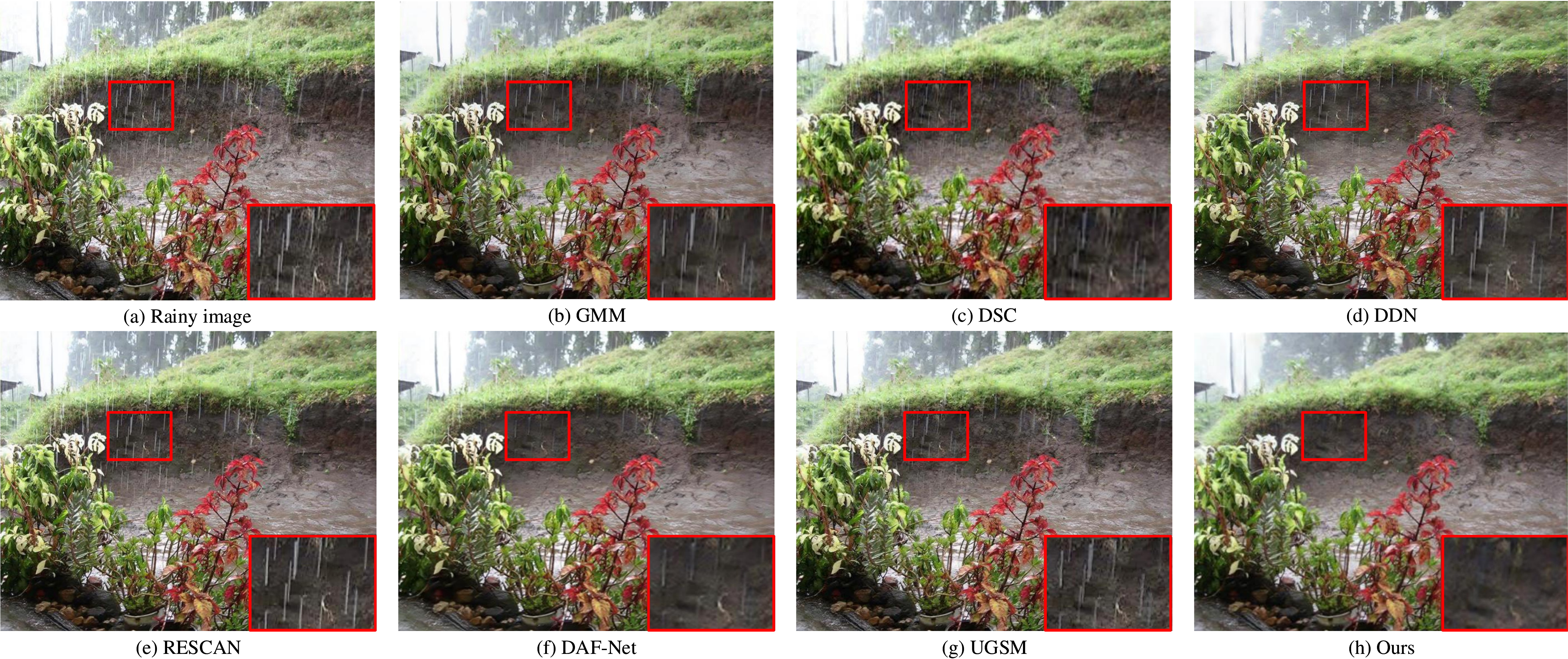}
	\caption{Image deraining results tested in real-world dataset. From (a)-(h): (a) the rainy image Plant, and the deraining results of (b) GMM \cite{Gaussian_mixture_model}, (c) DSC \cite{DSC}, (d) DDN \cite{detail_layer},
		(e) RESCAN \cite{rescan}, (f) DAF-Net \cite{hu2019}, (g) UGSM \cite{AMM}, and (h) our DRD-Net respectively. }
	\label{fig:real_1}
\end{figure*}

\begin{figure*}[ht] \centering
	\includegraphics[width=1\linewidth]{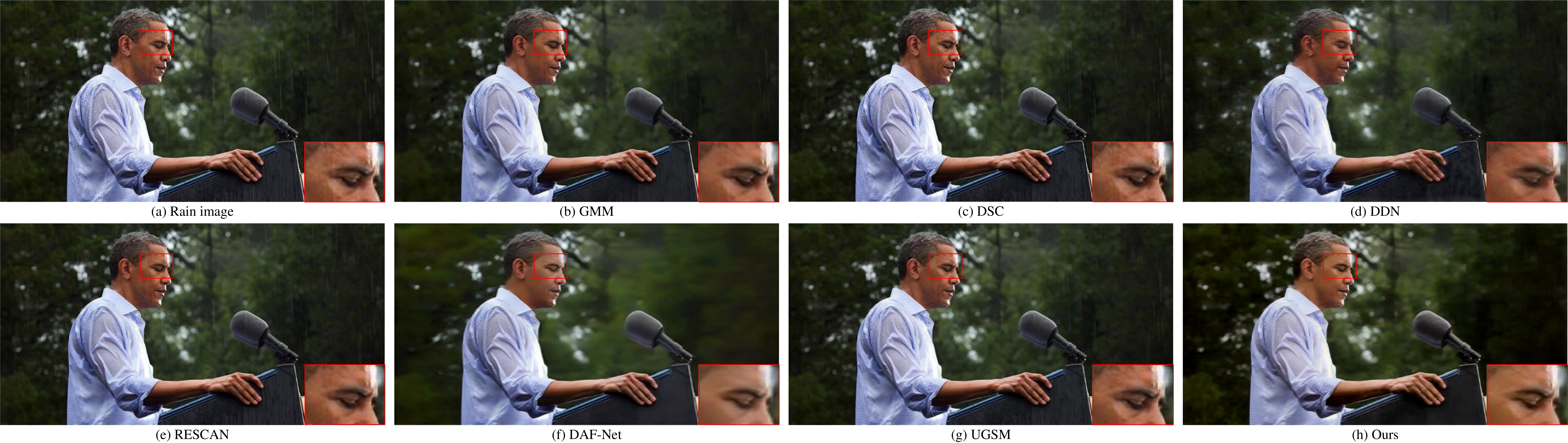}
	\caption{Image deraining results tested in real-world dataset. From (a)-(h): (a) the rainy image Obama, and the deraining results of (b) GMM \cite{Gaussian_mixture_model}, (c) DSC \cite{DSC}, (d) DDN \cite{detail_layer},
		(e) RESCAN \cite{rescan}, (f) DAF-Net \cite{hu2019}, (g) UGSM \cite{AMM}, and (h) our DRD-Net respectively.}
	\label{fig:real_2}
\end{figure*}

\begin{figure*}[ht] \centering
	\includegraphics[width=1\linewidth]{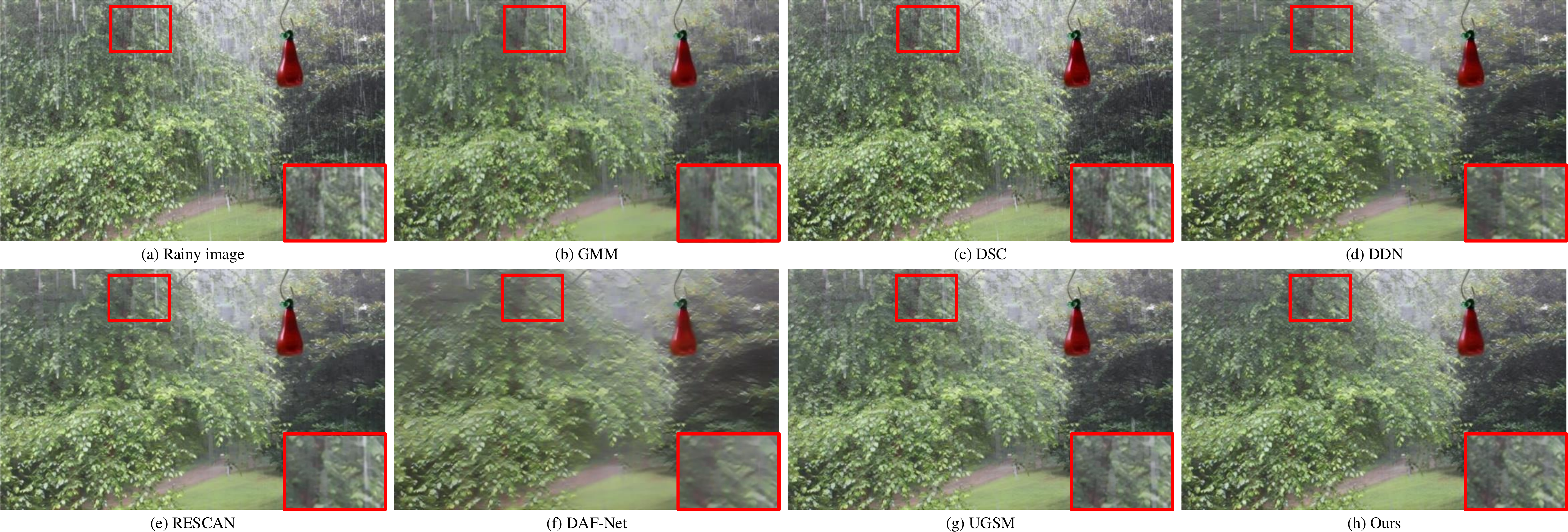}
	\caption{Image deraining results tested in real-world dataset. From (a)-(h): (a) the rainy image Trees, and the deraining results of (b) GMM \cite{Gaussian_mixture_model}, (c) DSC \cite{DSC}, (d) DDN \cite{detail_layer},
		(e) RESCAN \cite{rescan}, (f) DAF-Net \cite{hu2019}, (g) UGSM \cite{AMM}, and (h) our DRD-Net respectively. }
	\label{fig:real-world_45}
\end{figure*}

\begin{figure*}[ht] \centering
	\includegraphics[width=1\linewidth]{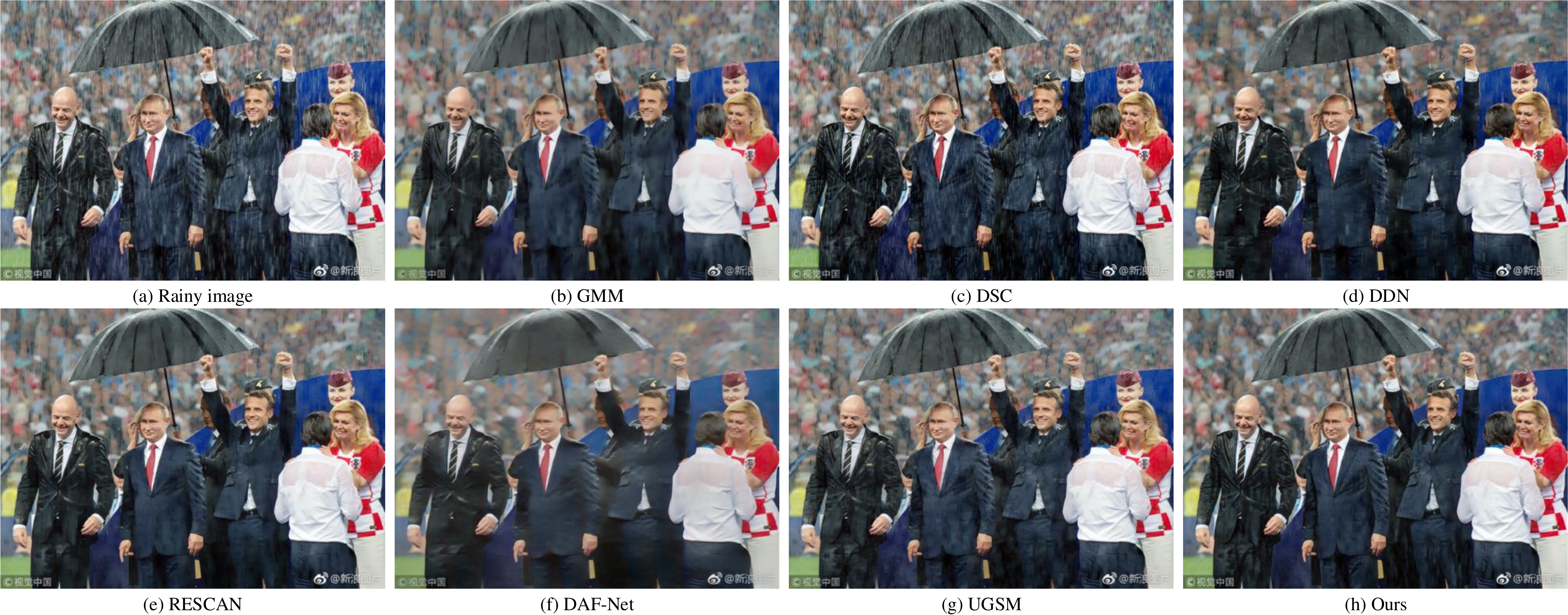}
	\caption{Image deraining results tested in real-world dataset. From (a)-(h): (a) the rainy image Putin, and the deraining results of (b) GMM \cite{Gaussian_mixture_model}, (c) DSC \cite{DSC}, (d) DDN \cite{detail_layer},
		(e) RESCAN \cite{rescan}, (f) DAF-Net \cite{hu2019}, (g) UGSM \cite{AMM}, and (h) our DRD-Net respectively. }
	\label{fig:real-world_100}
\end{figure*}

\begin{figure*}[ht] \centering
	\includegraphics[width=1\linewidth]{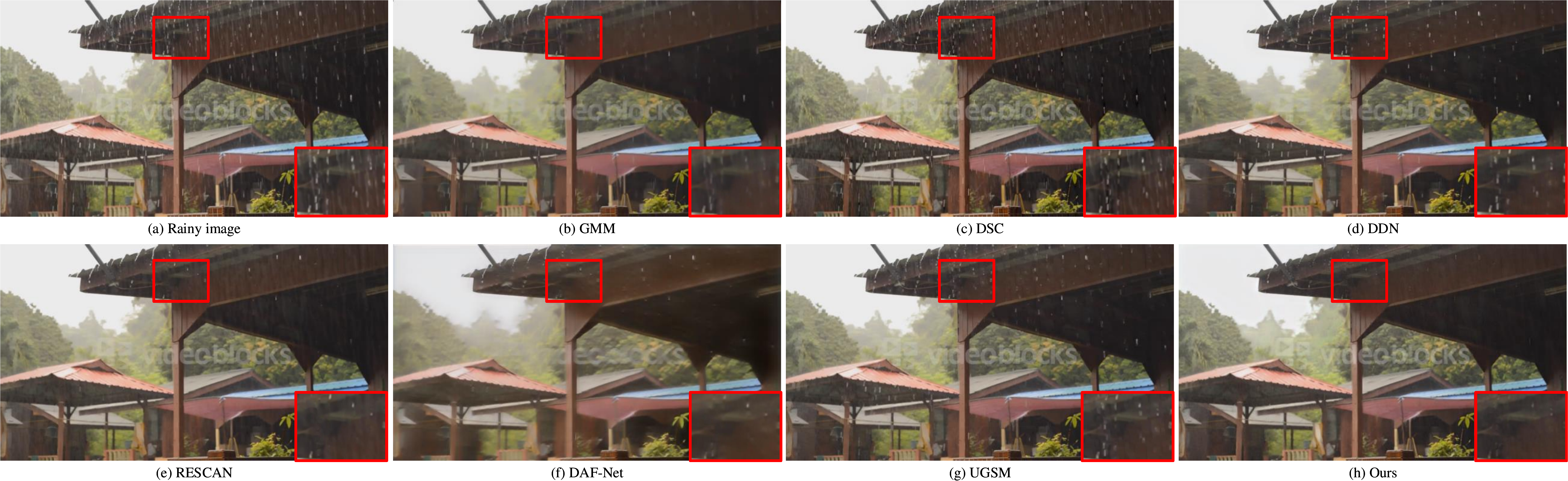}
	\caption{Image deraining results tested in real-world dataset. From (a)-(h): (a) the rainy image House, and the deraining results of (b) GMM \cite{Gaussian_mixture_model}, (c) DSC \cite{DSC}, (d) DDN \cite{detail_layer},
		(e) RESCAN \cite{rescan}, (f) DAF-Net \cite{hu2019}, (g) UGSM \cite{AMM}, and (h) our DRD-Net respectively. }
	\label{fig:real-world_131}
\end{figure*}

\begin{figure*}[thp] \centering
	\includegraphics[width=1\linewidth]{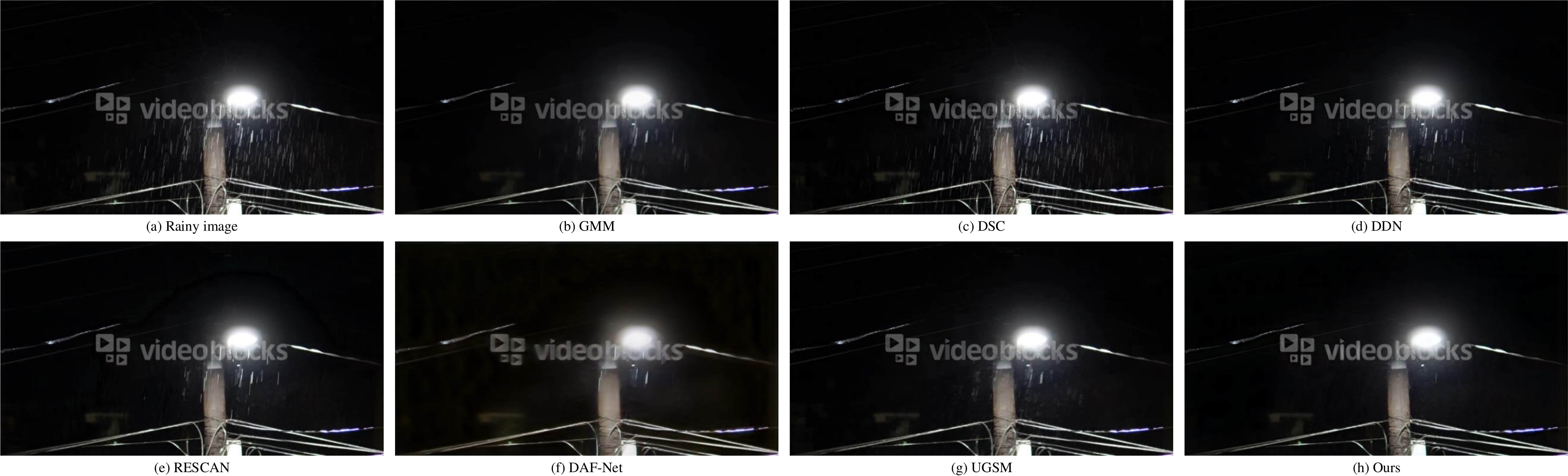}
	\caption{Image deraining results tested in real-world dataset. From (a)-(h): (a) the rainy image Street Lamp, and the deraining results of (b) GMM \cite{Gaussian_mixture_model}, (c) DSC \cite{DSC}, (d) DDN \cite{detail_layer},
		(e) RESCAN \cite{rescan}, (f) DAF-Net \cite{hu2019}, (g) UGSM \cite{AMM}, and (h) our DRD-Net respectively. }
	\label{fig:real-world_138}
\end{figure*}

\begin{figure*}[t] \centering
	\includegraphics[width=1\linewidth]{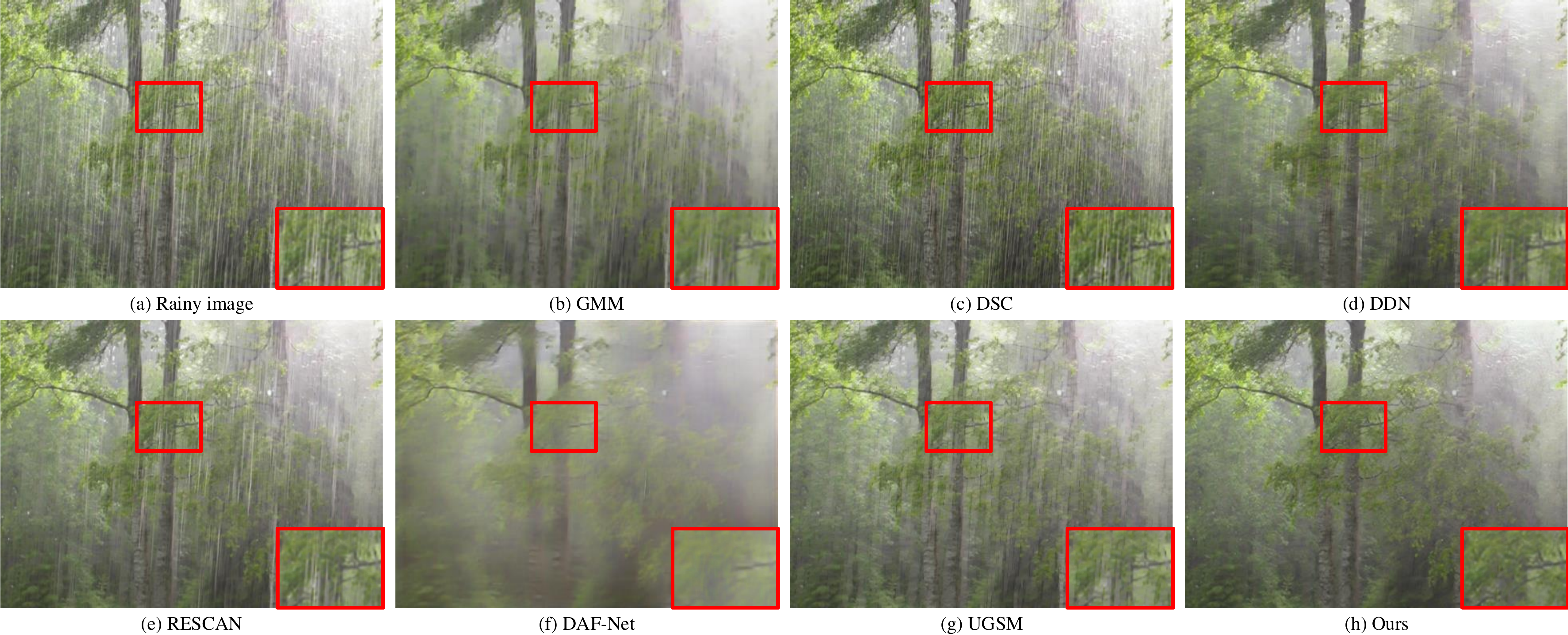}
	\caption{Image deraining results tested in real-world dataset. From (a)-(h): (a) the rainy image Rainforest, and the deraining results of (b) GMM \cite{Gaussian_mixture_model}, (c) DSC \cite{DSC}, (d) DDN \cite{detail_layer},
		(e) RESCAN \cite{rescan}, (f) DAF-Net \cite{hu2019}, (g) UGSM \cite{AMM}, and our (h) DRD-Net respectively. }
	\label{fig:real-world_37}
\end{figure*}

\subsection{Test on Real-world Datasets}

\par In order to validate the practicability of our DRD-Net, we visually evaluate its performance on a series of real-world rainy images in Figs. \ref{fig:real_1}-\ref{fig:real-world_37}. The DRD-Net can effectively remove the real-world rain streaks from the images while preserving their details, but the other approaches somewhat tend to over-smooth the images.

\subsection{Discussion and Ablation Study}

\par To explore the effectiveness of our DRD-Net, it is necessary to decompose its full scheme into different parts and even replace the network architecture for the ablation study.

\begin{itemize}
	\item \textbf{BL:} Baseline (BL) indicates the residual network without the SE operation, which learns a function that maps the rainy images to the rain streaks.
	
	\item \textbf{BL+SE:} Adding the SE operation to the baseline.
	
	\item \textbf{BL+SE+DB:} Employing two sub-networks for image deraining. One network is the rain residual network (BL+SE), and the another is detail repair network based on the direct block (DB), which is shown in Fig. \ref{fig:different connect}(a).
	
	\item \textbf{BL+SE+RB:} Different from the scheme of BL+SE+DB, the direct block (DB) is replaced with residual block (RB) in the detail repair network.
	
	\item \textbf{BL+SE+SDCAB:} Our proposed detail-recovery image deraining network (DRD-Net), which comprises the rain residual network (BL+SE) and the detail repair network based on the proposed structure detail context aggregation block (SDCAB).
	
\end{itemize}

\par \textbf{Analysis on SE:} Considering the rain streaks in the heavy rain images have different distributions of directions, colors and shapes \cite{rescan}, we introduce the SE operation into the rain residual network to determine which kind of the rain streaks are much more important. The results in Fig. \ref{fig:se} have demonstrated the effectiveness of the SE operation. Moreover, we remove the SE operation from the network and show the results in Table \ref{Tab:ablation_1}. It is found that the performance of deraining without the SE operation suffers from slight degradations. This certifies the necessity of the SE operation from another side.
\begin{table*}[htbp]
	\centering
	\caption{Quantitative comparison between our DRD-Net and other network architectures.}
	\label{Tab:ablation_1}
	\begin{threeparttable}
		\centering
		\setlength{\tabcolsep}{5.1mm}{
			\begin{tabular}{cccccccc}
				\toprule
				Dataset & Metrics & BL & BL+SE & BL+SE+DB & BL+SE+RB & BL+SE+SDCAB (Ours) \cr
				\toprule
				\multirow{2}{*}{Rain200L}
				& PSNR & 35.57 & 36.17 & 36.89 & \underline{37.04} & \textbf{37.15}  \cr
				& SSIM & 0.9759 & 0.9778 & 0.9792 & \underline{0.9860} & \textbf{0.9873}  \cr
				\multirow{2}{*}{Rain200H}
				& PSNR & 26.20 & 26.49 & \underline{27.16} & 27.01 & \textbf{28.16}  \cr
				& SSIM & 0.8245 & 0.8473 & \underline{0.9158} & 0.9061 & \textbf{0.9201}  \cr
				\multirow{2}{*}{Rain800}
				& PSNR & 25.83 & 26.04 & 26.09 & \underline{26.12} & \textbf{26.32}  \cr
				& SSIM & 0.8093 & 0.8181 & 0.8903 & \underline{0.8966} & \textbf{0.9018}  \cr
				\bottomrule
			\end{tabular}
		}
	\end{threeparttable}
\end{table*}

\par \textbf{Analysis on SDCAB:} The receptive field plays a significant role in image deraining \cite{rescan}. We propose a new connection style, named SDCAB, which has a large receptive field to build a detail repair network. In order to evaluate the effectiveness of the SDCAB, we compare our network with other connection style blocks, including the direct block (DB), the residual block (RB) which has been used in DDN \cite{detail_layer}. For fair comparison, we replace the SDCAB with DB and RB respectively. The results are shown in Table \ref{Tab:ablation_1}. The full scheme of BL+SE+SDCAB outperforms other architectures in the three datasets, which certifies that the SDCAB is essential to detail-recovery image deraining.

\par \textbf{Running Time:} We compare the running time of our method with different approaches on the dataset of Rain200H as shown in Table. \ref{Tab:time}. It is observed that our method is not the fastest one, but its performance is still acceptable.

\par \textbf{Ablation study:} The key components of our proposed network can be evaluated by an ablation study. We have discussed the effects of the number of feature maps and SDCAB or the rain residual blocks (RRB), which are shown in Table. \ref{Tab:ablation_2}.

\begin{table}[htbp]
	\centering
	\caption{Averaged time (in seconds) and performance of different methods in the dataset of Rain200H. }
	\label{Tab:time}	
	\begin{threeparttable}
		\centering
		\setlength{\tabcolsep}{1.2mm}{
			\begin{tabular}{cccccccc}
				\toprule
				Metrics & GMM & DSC & DDN & RESCAN & DAF-Net & Ours \cr
				\toprule
				PSNR & 13.04 & 13.17 & 24.64 & 26.60 & 24.65 & 28.16 \cr
				Avg time & 331.4s & 92.9s & 0.03s & 0.25s & 0.52s & 0.54s \cr
				\bottomrule
			\end{tabular}
		}
	\end{threeparttable}
	
\end{table}
\begin{table}[htbp]
	\centering
	\caption{Ablation study on different settings of our method on synthetic dataset Rain200H. The M denotes the number of feature maps in our network and the D is the total depth of our network.}
	\label{Tab:ablation_2}
	\begin{threeparttable}
		\centering
		\setlength{\tabcolsep}{3mm}{
			\begin{tabular}{ccccc}
				\toprule
				& Metrics & M = 16 & M = 32 & M = 64  \cr
				\toprule
				\multirow{2}{*}{D = 8+3}
				& PSNR & 26.36 & 26.77 & 26.97 \cr
				& SSIM & 0.9085 & 0.9117 &0.9135 \cr
				\multirow{2}{*}{D = 12+3}
				& PSNR & 26.52 & 26.89 & 27.31  \cr
				& SSIM & 0.9092 & 0.9129 & 0.9152  \cr
				\multirow{2}{*}{D = 16+3}
				& PSNR & 26.93 & 27.61 & 28.16  \cr
				& SSIM & 0.9127 & 0.9183 & 0.9201  \cr
				\bottomrule
			\end{tabular}
		}
	\end{threeparttable}
\end{table}

\par
\begin{figure*}[htbp]
	\centering
	\subfigure[]{
		\includegraphics[width=2.27in, height=1.8in]{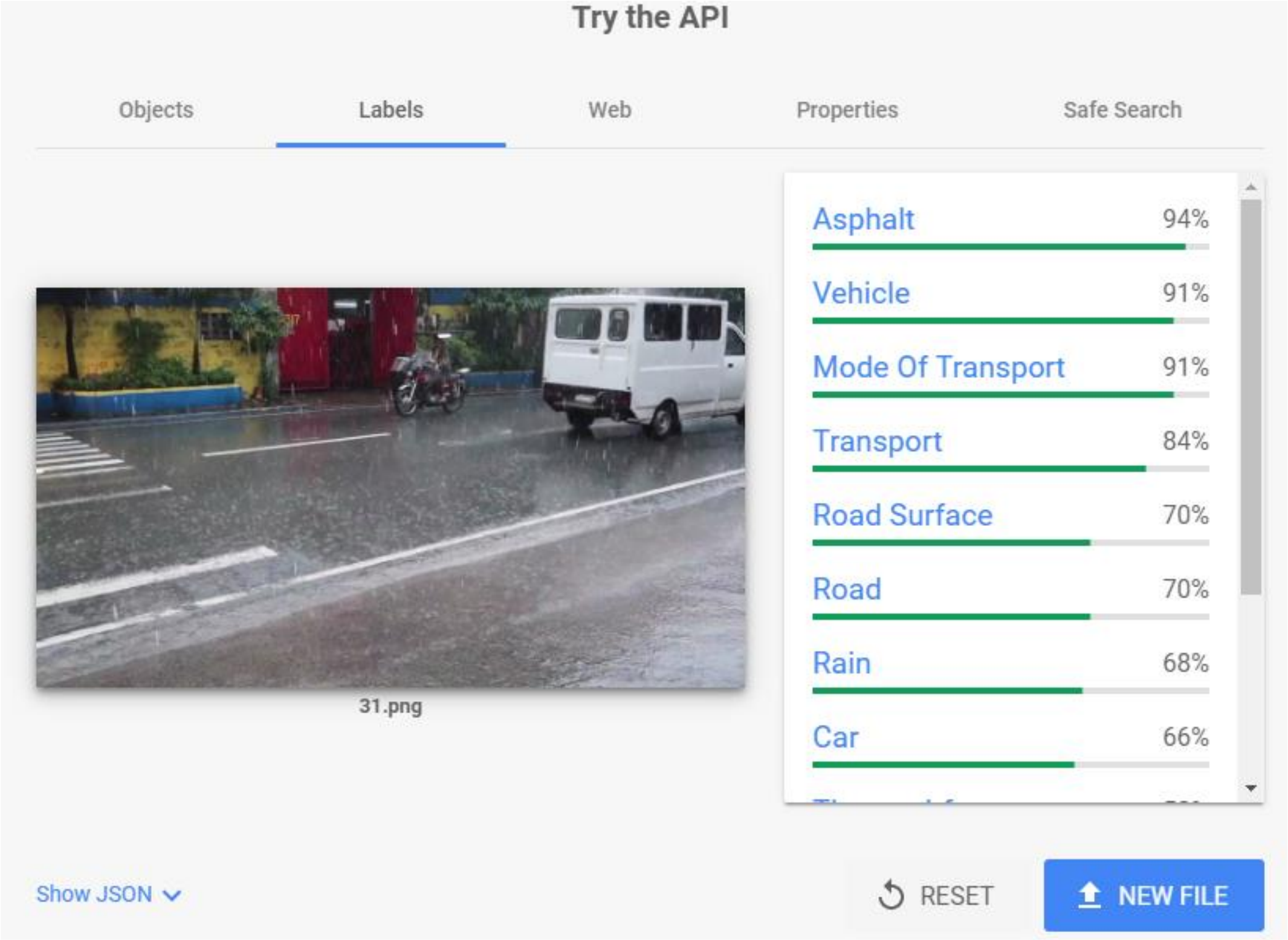}
	}
	\subfigure[]{
		\includegraphics[width=2.27in, height=1.8in]{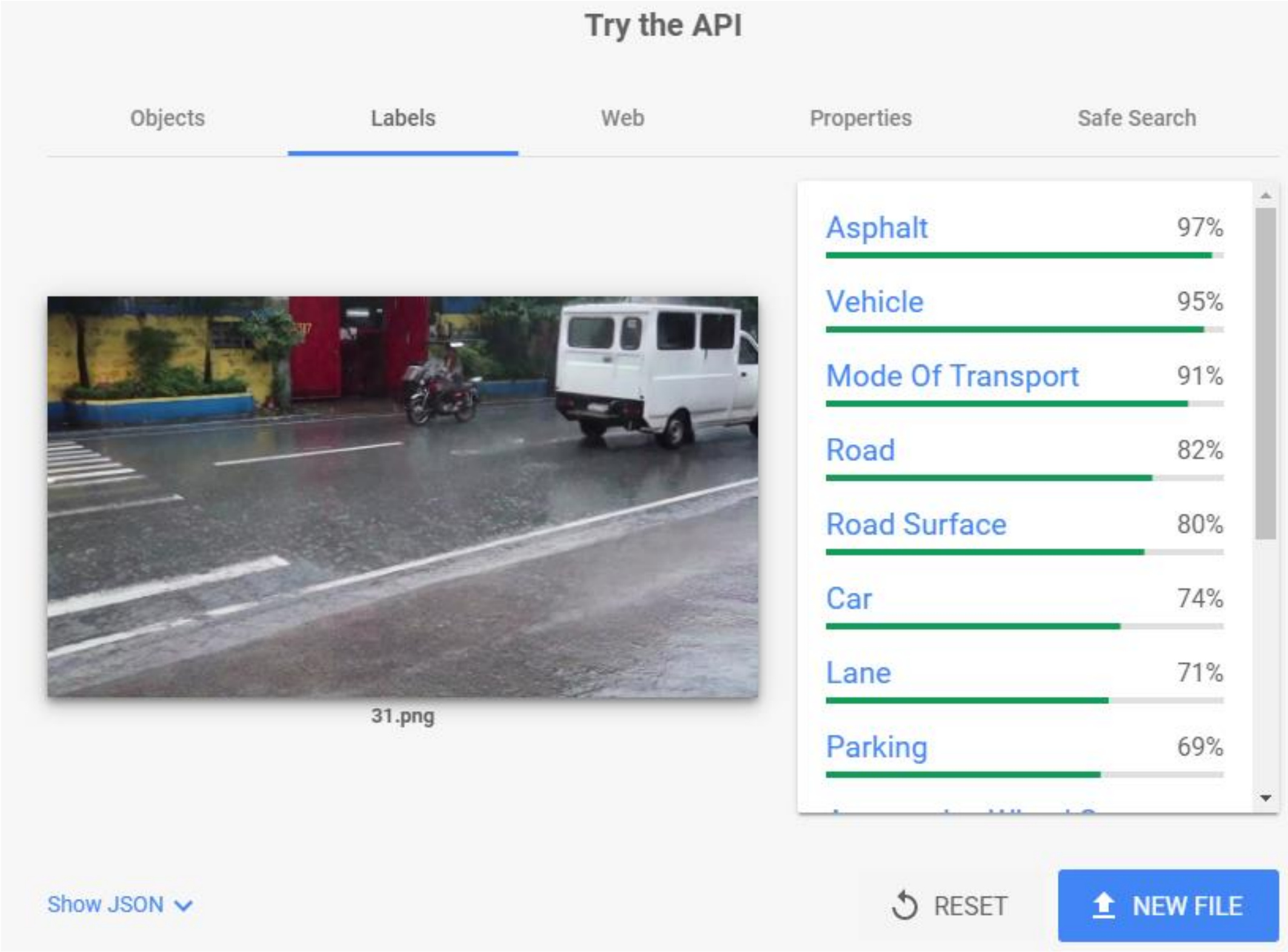}
	}
	\subfigure[]{
		\includegraphics[width=2.27in, height=1.8in]{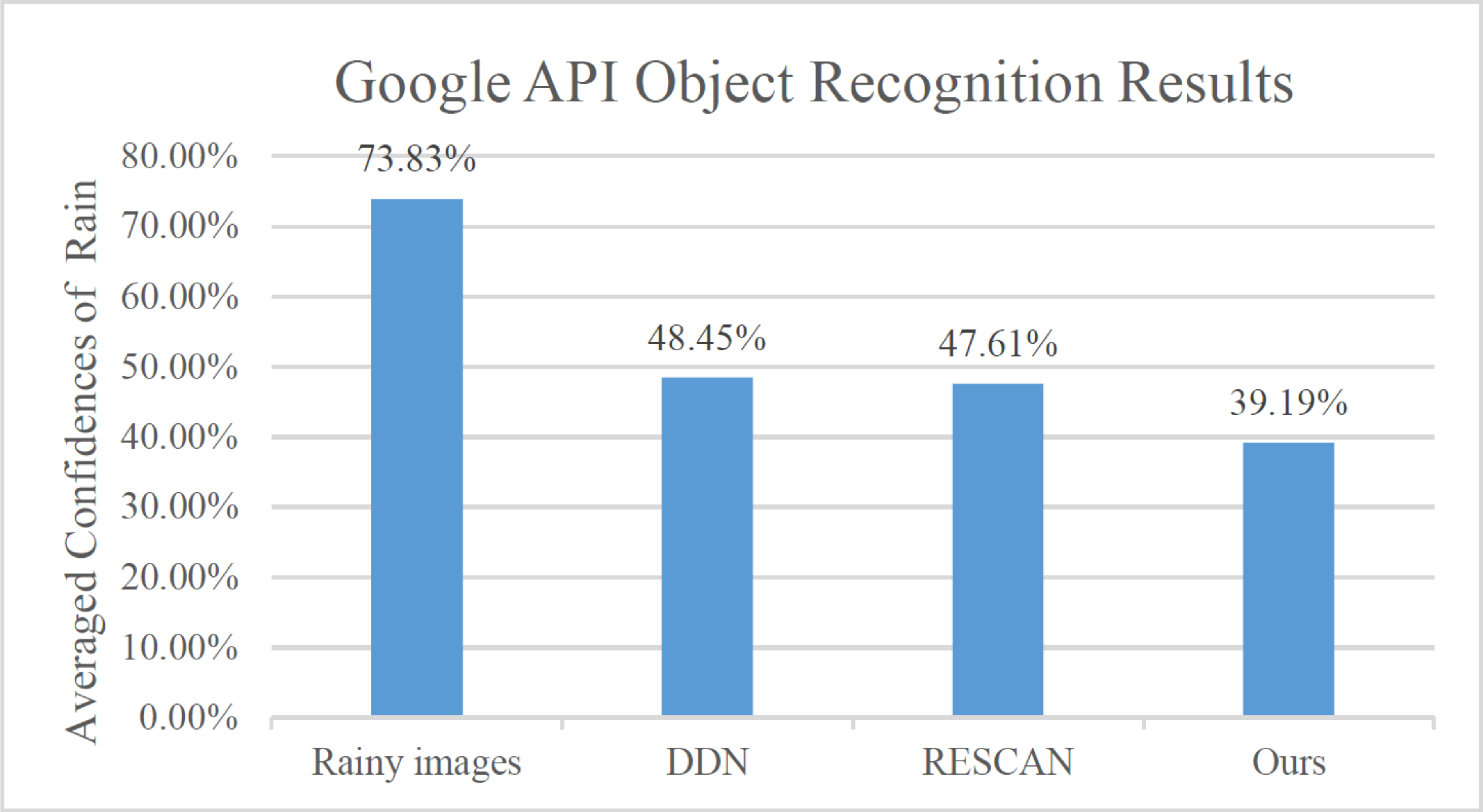}
	}
	\caption{The deraining results tested on the Google Vision API. From (a)-(c): (a) object recognition result in the real-world rainy image, (b) object recognition result after deraining by our DRD-Net, and (c) the averaged confidences in recognizing rain from 30 sets of the real-world rainy images and derained images of DDN \cite{detail_layer}, RESCAN \cite{rescan} and Our DRD-Net respectively. Note: zero confidence refers to a total failure in recognizing rain from a derained image by Google API.}
	\label{fig:google_api}
\end{figure*}

 \textbf{Application:} To demonstrate that our deraining method can benefit computer vision applications, we employ Google Vision API to evaluate the deraining results. One of the results is shown in Fig. \ref{fig:google_api} (a-b). It is observed that the Google API can recognize the rainy weather in the rainy image while it cannot recognize the rainy weather in the derained image. Furthermore, we use the Google API to test 30 sets of the real-world rainy images and derained images of our method and two baseline methods \cite{rescan, detail_layer}, and the result is shown in Fig. \ref{fig:google_api} (c). As one can see, after deraining, the confidences in recognizing rain from the images are significantly reduced.

\par \textbf{Limitation:} We use two sub-networks to deal with the problem of removing rain from single images. Although the proposed technique can simultaneously derain the image and preserve the details, it requires more parameters and takes a little more time to train the network.

\section{Conclusions}

\par We have presented an end-to-end network with two sub-networks for image deraining from single images.
One network is designed to remove the rain streaks from the rainy images, the other is proposed to find back the details to the derained image.
We propose the new structure detail context aggregation block (SDCAB) which has a large receptive field to obtain more spatial information. Moreover, qualitative and quantitative experiments indicate that our method outperforms the state-of-the-art learning-based and traditional approaches in terms of removing the rain streaks and recovering the image details.

\ifCLASSOPTIONcaptionsoff
  \newpage
\fi



%

%
%

\bibliographystyle{IEEEtran}

\bibliography{pcf}

\begin{IEEEbiography}[{\includegraphics[width=1in,height=1.25in,clip,keepaspectratio]{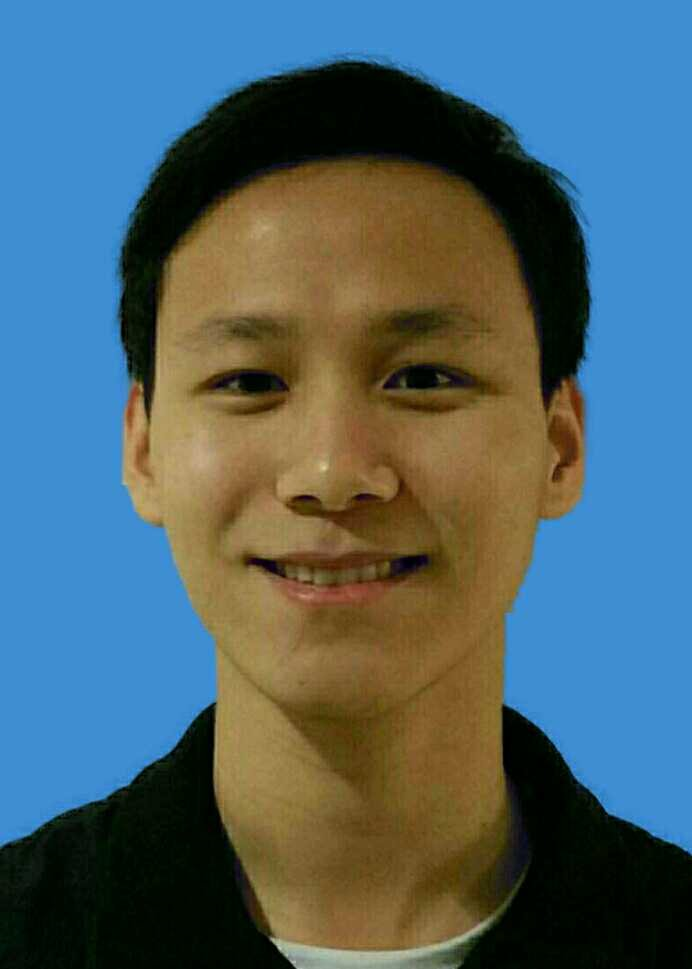}}]{Sen Deng}
	is now pursuing his PhD degree at the School of Computer Science and Technology, Nanjing University of Aeronautics and Astronautics (NUAA), China, from 2018. He received his Bachelor's degree in the University of Electronic Science and Technology of China. His research interests include deep learning, image processing and computer vision.
\end{IEEEbiography}

\begin{IEEEbiography}[{\includegraphics[width=1in,height=1.25in,clip,keepaspectratio]{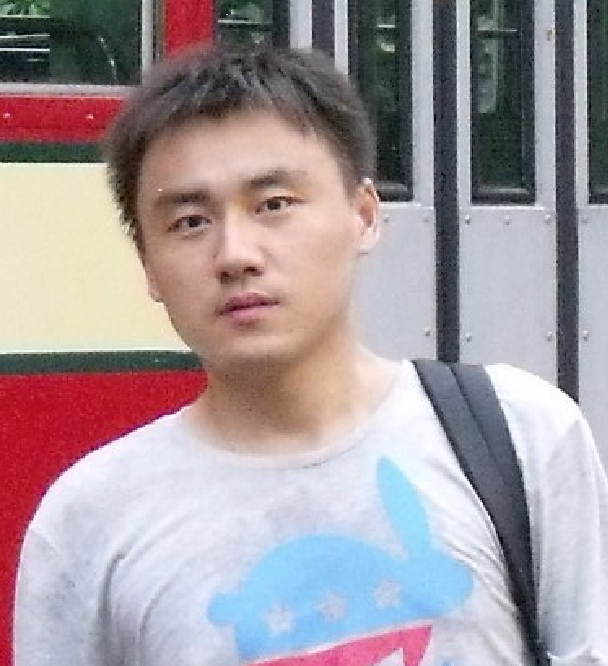}}]{Mingqiang Wei}
	received his Ph.D degree (2014) in Computer Science and Engineering from the Chinese University of Hong Kong (CUHK). He is an associate professor at the School of Computer Science and Technology, Nanjing University of Aeronautics and Astronautics (NUAA). Before joining NUAA, he served as an assistant professor at Hefei University of Technology, and a postdoctoral fellow at CUHK. His research interest is computer graphics with an emphasis on smart geometry processing.
\end{IEEEbiography}

\begin{IEEEbiography}[{\includegraphics[width=1in,height=1.25in,clip,keepaspectratio]{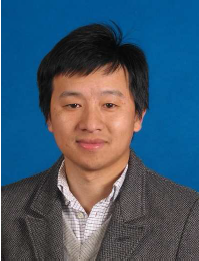}}]{Jun Wang}
	is currently a professor at Nanjing University of Aeronautics and Astronautics (NUAA), China. He received his Bachelor and PhD degrees in Computer-Aided Design from NUAA in 2002 and 2007 respectively. From 2008 to 2009, he conducted research as a postdoctoral scholar at the University of California, Davis. Subsequently, he worked as a research associate
	at the University of Wisconsin, Milwaukee for one year. From 2010 to 2013, he worked as a senior research engineer at Leica Geosystems Inc., USA. In 2013, he paid an academic visit to the Department of Mathematics at Harvard University. His research interests include geometry processing and geometric modeling, especially large-scale LiDAR point data capturing, management, processing and analysis.
\end{IEEEbiography}

\begin{IEEEbiography}[{\includegraphics[width=1in,height=1.25in,clip,keepaspectratio]{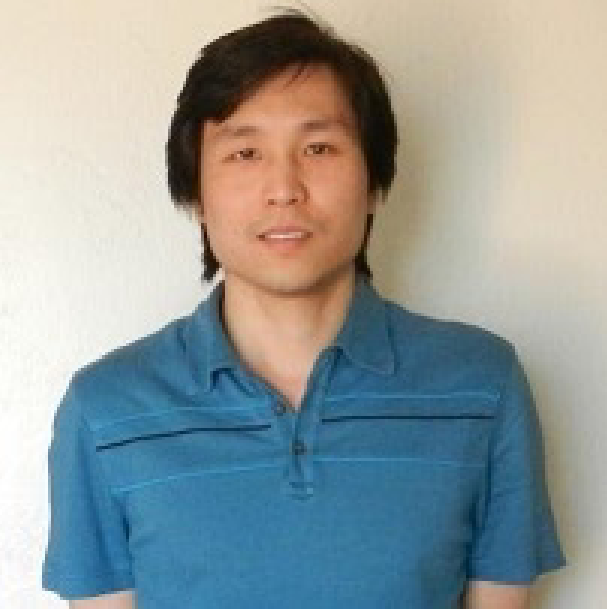}}]{Luming Liang}
	is a senior researcher at Microsoft. Before working at Microsoft, Luming was a software engineer and then a data \& applied scientist at Uber and Microsoft, respectively. He received his BSc. and MSc. in the School of Information Science and Engineering from Central South University, China, in 2005 and 2008, respectively and his PhD in the Department of Electrical Engineering and Computer Science from Colorado School of Mines, USA in 2014. His primary research interest is finding shape correspondences and image analysis.
\end{IEEEbiography}


\begin{IEEEbiography}[{\includegraphics[width=1in,height=1.25in,clip,keepaspectratio]{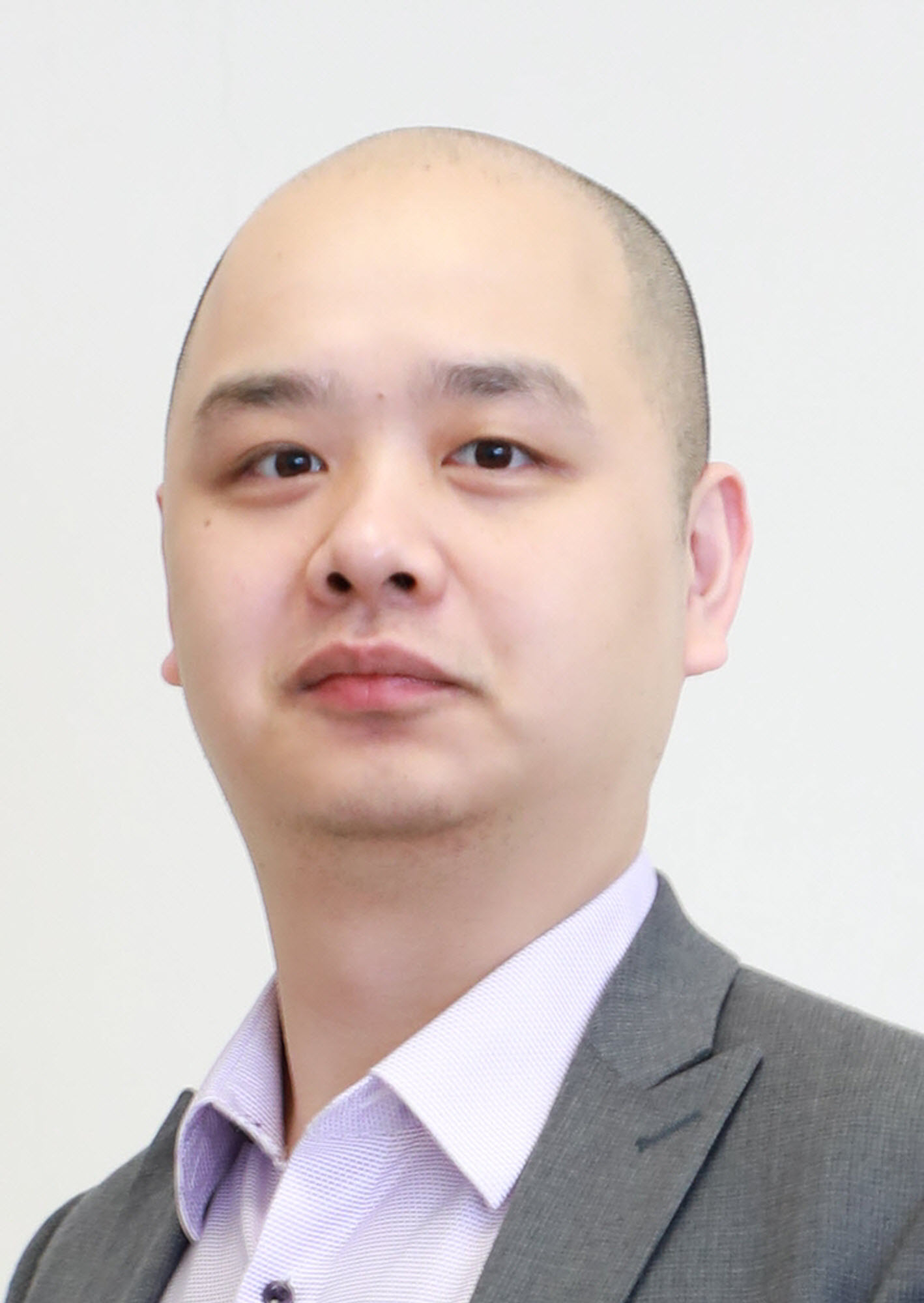}}]{Haoran Xie}
		is an Assistant Professor at The Education University of Hong Kong. He received his PhD in Computer Science from the City University of Hong Kong. His research interest includes artificial intelligence, big data, and educational technology. He has totally published 170 research publications including 71 journal articles (34 papers as rst or corresponding author), 40 LNCS/CCIS book chapters, 54 conference proceedings, 4 edited books and 1 patent. Among all 59 journal articles, there are 52 SCI/SSCI indexed and 12 SCOPUS indexed. He has obtained 10 research awards including the Golden Medal and the special award from International Invention Innovation Competition in Canada, theSilver Award from Geneva's Invention Expo, 2nd Prize Winner from Multimedia Grand Challenges of ACM Multimedia 2019, Outstanding Research Achievement Award from Asia Pacic Society for Computing and Information Technology, President's Award for Outstanding Performance in Research in EdUHK, Faculty's Research Output Award in EdUHK, and 3 best/excellent paper awards from international conferences including DASFAA 2017, ICBL 2016 and SECOP 2015. His proposed LSGAN published in IEEE TPAMI and ICCV, with more than 700 citations in two years, has been included in the computer vision course in Stanford University and implemented by Google TensorFlow. Dr. Xie has served as 9 guest editors in special issues of journals, editors of 4 books, organization committee chairs/members of more than 55 conferences, and reviewers of 28 international journals. He has successfully obtained 36 research grants including some external competitive research grants such as ITF, IDS, UGC T\&L. The totally amount of these grants are more than HK\$ 20 millions.
\end{IEEEbiography}

\begin{IEEEbiography}[{\includegraphics[width=1in,height=1.25in,clip,keepaspectratio]{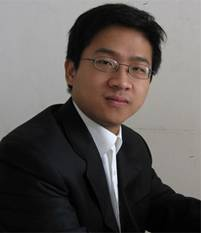}}]{Meng Wang}
	is a Professor with Hefei University of Technology, China. He received the B.E. degree and the Ph.D. degree in the Special Class for the Gifted Young from the Department of Electronic Engineering and Information Science, University of Science and Technology of China (USTC), Hefei, China, in 2003 and 2008, respectively. His current research interests include multimedia content analysis, computer vision, and pattern recognition. He has authored over 200 book chapters, journal papers, and conference papers in these areas. He was a recipient of the ACM SIGMM Rising Star Award 2014. He is an Associate Editor of IEEE TRANSACTIONS ON KNOWLEDGE AND DATA ENGINEERING, IEEE TRANSACTIONS ON CIRCUITS AND SYSTEMS FOR VIDEO TECHNOLOGY, and IEEE TRANSACTIONS ON NEURAL NETWORKS AND LEARNING SYSTEMS.
\end{IEEEbiography}

%

%

%
%
%




\end{document}